\documentclass[journal,twocolumn,10pt]{IEEEtran}

\ifCLASSINFOpdf
  \usepackage[pdftex]{graphicx}
  \usepackage{epstopdf}
  \graphicspath{{../Figures/}}
  \DeclareGraphicsExtensions{.eps}
\else
  \usepackage[dvips]{graphicx}
\fi

\usepackage{graphicx}
\usepackage{amssymb}
\usepackage{amsmath}
\usepackage{mathtools}
\usepackage{dsfont}
\usepackage{cite}
\usepackage{stfloats}
\usepackage{subfigure}
\usepackage{psfrag}
\usepackage[mathscr]{euscript}
\usepackage{algorithm, algorithmic}
\usepackage{acronym}  
\usepackage{booktabs}
\usepackage{bbm}

\usepackage{float}
\usepackage{balance}

\acrodef{CCDF}{complementary cumulative distribution function}
\acrodef{CF}{characteristic function}
\acrodef{PPP}{Poisson point processe}
\acrodef{RV}{random variable}
\acrodef{i.i.d.}{independent and identically distributed}
\acrodef{PDF}{probability distribution function}
\acrodef{CDF}{cumulative distribution function}
\acrodef{ch.f.}{characteristic function}
\acrodef{AWGN}{additive white Gaussian noise}
\acrodef{SNR}{signal-to-noise ratio}
\acrodef{LRT}{likelihood ratio test}
\acrodef{DRT}{distance ratio test}
\acrodef{GLRT}{generalized likelihood ratio test}
\acrodef{CRLB}{Cram\'{e}r-Rao lower bound}
\acrodef{CRB}{Cram\'{e}r-Rao bound}
\acrodef{ZZLB}{Ziv-Zakai lower bound}
\acrodef{ZZB}{Ziv-Zakai bound}
\acrodef{LOS}{line-of-sight}
\acrodef{ToF}{time-of-flight}
\acrodef{NLOS}{non-line-of-sight}
\acrodef{GDOP}{geometric dilution of precision}
\acrodef{GPS}{Global Positioning System}
\acrodef{FIM}{Fisher information matrix}
\acrodef{PEB}{position error bound}
\acrodef{SPEB}{squared position error bound}
\acrodef{TOA}{time-of-arrival}
\acrodef{TOF}{time-of-flight}
\acrodef{WSN}{wireless sensor network}
\acrodef{MAC}{medium access control}
\acrodef{RSS}{received signal strength}
\acrodef{WAF}{wall attenuation factor}
\acrodef{TDOA}{time difference-of-arrival}
\acrodef{RF}{radiofrequency}
\acrodef{RTT}{round-trip time}
\acrodef{AOA}{angle-of-arrival}
\acrodef{MF}{matched filter}
\acrodef{ED}{energy detector}
\acrodef{ML}{maximum likelihood}
\acrodef{MSE}{mean-square error}
\acrodef{RMSE}{root-mean-square error}
\acrodef{LEO}{localization error outage}
\acrodef{ppm}{part-per-million}
\acrodef{ACK}{acknowledge}
\acrodef{UWB}{Ultrawide bandwidth}
\acrodef{TNR}{threshold-to-noise ratio}
\acrodef{LS}{least squares}
\acrodef{IR-UWB}{impulse radio UWB}
\acrodef{FCC}{Federal Communications Commission}
\acrodef{TH}{time-hopping}
\acrodef{PPM}{pulse position modulation}
\acrodef{MUI}{multi-user interference}
\acrodef{PDP}{power delay profile}
\acrodef{BPZF}{band-pass zonal filter}
\acrodef{SIR}{signal-to-interference ratio}
\acrodef{SINR}{signal-to-interference-plus-noise ratio}
\acrodef{RFID}{radio frequency identification}
\acrodef{WPAN}{wireless personal area network}
\acrodef{WWB}{Weiss-Weinstein bound}
\acrodef{DP}{direct path}
\acrodef{MF}{matched filter}
\acrodef{MMSE}{minimum-mean-square-error}
\acrodef{SBS}{serial backward search}
\acrodef{SBSMC}{serial backward search for multiple clusters}
\acrodef{NBI}{narrowband interference}
\acrodef{WBI}{wideband interference}
\acrodef{INR}{interference-to-noise ratio}
\acrodef{CR}{channel response}
\acrodef{CIR}{channel impulse response}
\acrodef{CR}{channel  response}
\acrodef{RADAR}{radar}
\acrodef{MUR}{Multistatic radar}
\acrodef{JBSF}{jump back and search forward}
\acrodef{HDSA}{high-definition situation-aware}
\acrodef{RRC}{root raised cosine}
\acrodef{ST}{simple thresholding}
\acrodef{BTB}{Bellini-Tartara bound}
\acrodef{P-Max}{$P$-Max}  
\acrodef{MIMO}{multiple-input multiple-output}
\acrodef{MAP}{maximum a posteriori}
\acrodef{FG}{factor graph}
\acrodef{OP}{outage probability}
\acrodef{WED}{wall extra delay}
\acrodef{RMS}{root mean square}
\acrodef{SPAWN}{sum-product algorithm over a wireless network}
\acrodef{MDD}{minimum distance distribution}
\acrodef{MAP}{maximum a posteriori probability}
\acrodef{SAP}{small cell access point}
\acrodef{UE}{user equipment}
\acrodef{MBS}{macro cell base station}
\acrodef{UER}{\ac{UE} Relay}
\acrodef{D2D}{device-to-device}
\acrodef{MBS}{macro base station}
\acrodef{CSI}{channel state information}
\acrodef{OGR}{outage guard region}
\acrodef{FUR}{feasible UER region}
\acrodef{EHR}{energy harvesting region}
\acrodef{EH}{energy harvesting}
\acrodef{D2D-EHSN}{D2D communication provided \ac{EH} small cell network}
\acrodef{D2D-EHHN}{D2D communication provided \ac{EH} heterogeneous network}
\acrodef{3GPP}{3rd Generation Partnership Project}
\acrodef{BS}{base station}
\acrodef{DF}{decode and forward}
\acrodef{CCDF}{complementary cumulative distribution function}
\acrodef{ZF}{zero forcing}
\acrodef{RZF}{regularized zero forcing}
\acrodef{WLLN}{weak law of large number}
\acrodef{SLLN}{strong law of large numbers}
\acrodef{TDD}{Time-division duplex}
\acrodef{EE}{energy efficiency} 
\acrodef{HetNet}{heterogeneous network} 
\acrodef{SCP}{Single Cell Processing}
\acrodef{CBF}{Coordinated Beamforming}
\usepackage{color}
\usepackage{dsfont}
\usepackage{bbm}

\def\PVT{P_{\mathrm{ut}}}

\DeclareMathAlphabet{\mathsf}{OML}{cmbr}{m}{it}

\newtheorem{theorem}{\bf Theorem}
\newtheorem{lemma}{\bf Lemma}
\newtheorem{corollary}{\bf Corollary}

\newtheorem{remark}{\bf Remark}

\newtheorem{assumption}{\bf Assumption}

\newcommand{\bd}{\begin{description}}
\newcommand{\ed}{\end{description}}
\newcommand{\be}{\begin{enumerate}}
\newcommand{\ee}{\end{enumerate}}
\newcommand{\bi}{\begin{itemize}}
\newcommand{\ei}{\end{itemize}}
\newcommand{\bl}{\begin{list}}
\newcommand{\el}{\end{list}}
\newcommand{\bt}{\begin{tabbing}}
\newcommand{\et}{\end{tabbing}}

\newcommand{\paperTitle}{ Scheduling Policies for Federated Learning in Wireless Networks }

\begin{document}

{
\title{\paperTitle}

\author{

      Howard~H.~Yang, \textit{Member, IEEE},
      Zuozhu~Liu, \textit{Student Member, IEEE},\\
      Tony~Q.~S.~Quek, \textit{Fellow, IEEE},
      and H.~Vincent~Poor, \textit{Fellow, IEEE}


\thanks{ 
H.~H.~Yang and T.~Q.~S.~Quek are with the Information System Technology and Design Pillar, Singapore University of Technology and Design (e-mail: \{howard\_yang, tonyquek\}@sutd.edu.sg).

Z.~Liu is with the Department of Statistics and Applied Probability, National University of Singapore (e-mail: staliuz@nus.edu.sg).

H.~V.~Poor is with the Department of Electrical Engineering, Princeton University, Princeton, NJ 08544 USA (e-mail: poor@princeton.edu).
}
}
\maketitle
\acresetall
\thispagestyle{empty}
\begin{abstract}
Motivated by the increasing computational capacity of wireless user equipments (UEs), e.g., smart phones, tablets, or vehicles, as well as the increasing concerns about sharing private data, a new machine learning model has emerged, namely federated learning (FL), that allows a decoupling of data acquisition and computation at the central unit.
Unlike centralized learning taking place in a data center, FL usually operates in a wireless edge network where the communication medium is resource-constrained and unreliable.
Due to limited bandwidth, only a portion of UEs can be scheduled for updates at each iteration. Due to the shared nature of the wireless medium, transmissions are subjected to interference and are not guaranteed.
The performance of FL system in such a setting is not well understood.
In this paper, an analytical model is developed to characterize the performance of FL in wireless networks.
Particularly, tractable expressions are derived for the convergence rate of FL in a wireless setting, accounting for effects from both scheduling schemes and inter-cell interference.
Using the developed analysis, the effectiveness of three different scheduling policies, i.e., random scheduling (RS), round robin (RR), and proportional fair (PF), are compared in terms of  FL convergence rate.
It is shown that running FL with PF outperforms RS and RR if the network is operating under a high signal-to-interference-plus-noise ratio (SINR) threshold, while RR is more preferable when the SINR threshold is low.
Moreover, the FL convergence rate decreases rapidly as the SINR threshold increases, thus confirming the importance of compression and quantization of the update parameters.
The analysis also reveals a trade-off between the number of scheduled UEs and subchannel bandwidth under a fixed amount of available spectrum.
\end{abstract}
\acresetall

\section{Introduction}\label{sec:intro}
Next-generation computing networks will encounter a paradigm shift from a conventional cloud computing setting, which aggregates computational resources in a data center, to edge computing systems which largely deploy computational power to the network edges to meet the needs of applications that demand very high bandwidth and low latency, as well as supporting resource-constrained nodes reachable only over unreliable network connections \cite{MaoYouZha:17,YanQue:17,DinTanLa:17,LeeLeeQue:19}.
Along with the burgeoning development of machine learning, it is expected that by leveraging computing capability in the edge nodes, usually access points (APs), future networks will be able to utilize local data to conduct intelligent inference and control on many activities, e.g., learning activities of mobile phone users, predicting health events from wearable devices, or detecting burglaries within smart homes \cite{ZhuLiuDu:18,ParSamBen:18}.
Due to the sheer volume of data generated, as well as the growing capability of computational power and the increasing concerns about sharing private data at end-user devices, it becomes more attractive to perform learning directly on user equipments (UEs) as opposed to sending raw data to an AP.
To this end, a new machine learning model has emerged, namely federated learning (FL), that allows decoupling of data acquisition and computation at the central unit \cite{KonMcMRam:15,KonMcMBren:16,MaMMooRam:16}.
Specifically, as illustrated by Fig.~\ref{fig:FL_WN}, an FL system optimizes a global model by repeating the following processes:
$i$) the UEs perform local computing with their own data to minimize a predefined empirical risk function and update the trained weights to the AP,
$ii$) the AP collects the updates from UEs and consults the FL unit to produce an improved global model, and $iii$) output from the FL model is redistributed to the UEs and the UEs conduct further local training by using the global model as a reference.
In this fashion, the global unit, i.e., the AP, is able to train a statistical model from the data stored on a swarm of end devices, i.e., the UEs, without sacrificing their privacy.
As such, the FL touts the trial as having smarter models, lower latency, and less power consumption, all while ensuring privacy. These properties identify the FL as one of the most promising technologies of future intelligent networks.

Nonetheless, to make FL possible, one needs to tackle new challenges that require a fundamental departure from the standard methods designed for distributed optimization \cite{KonMcMRam:15}. In particular, different from traditional machine learning systems, where an algorithm runs on a large data set partitioned homogeneously across multiple servers in the cloud, FL is usually trained from a large non-i.i.d., and often unbalanced, data set generated by distinct distributions across different UEs.
Just as crucial is what could happen at the parameter update stage: While an iterative algorithm running on FL requires very low latency and high throughput connection between computing units, the AP generally needs to link a vast number of UEs through a resource-constrained spectrum and thus can allow only a limited number of UEs to send their trained weights via unreliable channels for global aggregation.
These challenges make issues such as stragglers and fault tolerance for FL significantly more important than for the conventional training in data centers.
To deliver a successful deployment of FL, network operators need to adopt new tools and a new way of thinking: model development and training with no direct access to the raw data, with communication cost as a limiting factor \cite{NisYon:18,WanHanWan:18}.
\begin{figure}[t!]
  \centering{}

    {\includegraphics[width=0.95\columnwidth]{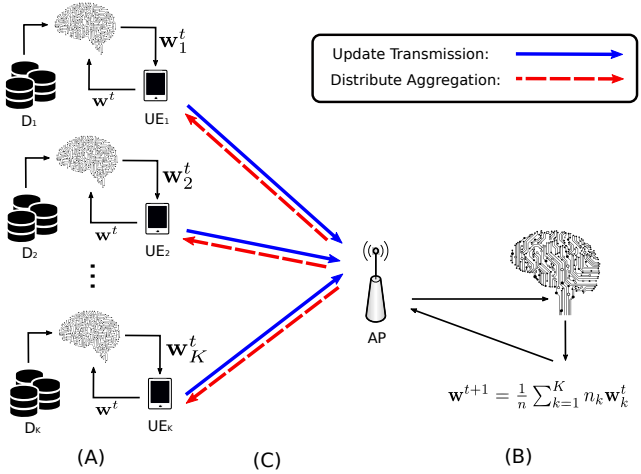}}

  \caption{ An illustration of the federated learning process: (A) each UE computes an individual update based on its locally stored data, (B) the AP aggregates the updates received from UEs to build a new global model, (C) the new model is sent back to the UEs, and the process is repeated. }
  \label{fig:FL_WN}
\end{figure}

In response, considerable research has been carried out, which can be mainly categorized into two directions: algorithmic and communication.
From an algorithmic perspective, the idea is to reduce the overhead in the update uploading phase to make the model training communication efficient,
where typical methods range from reducing the communication bandwidth by only updating the UEs with significant training improvement \cite{CheGiaSun:18}, compressing the gradient vectors via quantization \cite{AjiHea:17}, or adopting a momentum method in the sparse update to accelerate the training process \cite{LinHanMao:18}.
Recognizing that the unique properties of the wireless channel are not fully explored, another series of studies have followed up from the communication perspective.
Particularly, when the amount of training time is limited, solutions are taken by adapting the number of locally computing steps to the variance of the global gradient \cite{WanTuoSal:18,WanTuoSal:19JSAC,WanHanWan:18}, or scheduling the maximum number of UEs in a given time frame \cite{NisYon:18}.
When spectral resources become the communication bottleneck, there are new methods exploiting the compute-over-air mechanism and arrive at
a jointly decode-and-average scheme at the edge computing unit \cite{ZhuWanHua:18,YanJiaShi:18}.
Moreover, if perfect channel state information (CSI) is not available at the receiver, the trade-off between delay and number of users selected for parameter updating has also been investigated \cite{HaZhaSim:19}.
Among the prior work, the setup of communication is assumed in the single-cell scenario where received signals are affected only by the additive noise and thus can be correctly decoded upon each global aggregation. However, to fully realize the potential of federated learning, it is necessary to scale up the deployment across a large distributed network.
In this context, due to the shared nature of the wireless medium, communications are subjected to inter-cell interference and can encounter failure.
Additionally, since the spectral resources are generally limited, one needs to appropriately schedule the UEs for channel access upon each global update. To this end, for the successful delivery of FL in large-scale wireless networks, a complete understanding of its performance when operating under different scheduling schemes with unreliable communication links becomes essential.

\subsection{Approach and Summary of Contributions}
In this paper, we develop an analytical framework to study the impact of different scheduling policies on the performance of FL in large-scale wireless networks. Specifically, we model the AP deployment and UE locations as independent Poisson point processes (PPPs), where every UE possesses a private data set and each AP needs to collaboratively learn a statistical model with its associated UEs through FL.
Recognizing the potential inefficiency of the conventional FL training approach \cite{WanTuoSal:19JSAC}, we leverage methods from distributed coordinate descent \cite{MaKonJag:17} and propose an algorithm that decouples the global averaging at the AP and local computing at each UE, whereas the partial solutions from UEs constitutes a proximal step toward the global optimal that implicitly accelarates the convergence.
By leveraging tools from optimization theory and stochastic geometry \cite{YanQue:ICC19,YanQue:19,YanGerQue:16}, we derive tractable expressions for the FL convergence rate in a general setting that accounts for the employed scheduling policy and inter-cell interference that affects the data transmission phases.
Our main contributions are summarized below.
\begin{itemize}
  \item We propose an algorithm to train an FL model in the context of wireless networks. The algorithm is able to decompose a global statistical model into a number of local subproblems that can be efficiently solved using only the data set residing on each UE, and the solution of each local problem constitutes a proximal step toward the global optimum, which has the potential to accelerate the convergence rate. Moreover, the learning rate of each UE is set to be adjustable to changes in the communication environment.
  \item We develop a formal framework to analyze the convergence performance of FL algorithms run on wireless networks. Our analysis provides a tractable expression of the convergence rate, which takes into account the key features of a wireless communication system, including the transmission scheduling policy, small-scale fading, large-scale path loss, and inter-cell interference.
  \item We present the convergence rate of FL under three practical scheduling policies, i.e., random scheduling (RS), round robin (RR), and proportional fair (PF). We also analyze the convergence rate of FL in three special cases where $i$) only one UE can be scheduled upon each global aggregation, $ii$) the AP collects more updates by allowing multiple communications before each global aggregation, and $iii$) all UEs send out the trained weights without scheduling in every communication round.
  \item Through our analysis, we show that under high SINR threshold, running FL with PF outperforms RS and RR in terms of convergence rate, while RR is preferable when the SINR threshold is low.
      Moreover, for networks operating under very low SINR thresholds, sending trained weights without scheduling can achieve better FL convergence rate than any scheduling methods employed.
      The FL convergence rate is shown to decrease rapidly as the SINR threshold increases, thus confirming the importance of compression and quantization of the update parameters.
  \item Our analysis also reveals that under a fixed amount of available spectrum, there exists a trade-off between the number of scheduled UEs and subchannel bandwidth in the optimization of FL convergence rate, which allows further design options.
\end{itemize}

The remainder of this paper is organized as follows. We introduce the system model in Section II.
In Section III, we detail the local computing and parameter update process to run FL in wireless networks.
In Section IV, we analyze the convergence rate of federated learning under various scheduling policies.
We show the numerical results in Section V to compare the effectiveness of different scheduling methods and obtain design insights.
We conclude the paper in Section VI.

\begin{table}
\caption{Notation Summary
} \label{table:notation}
\begin{center}
\renewcommand{\arraystretch}{1.3}
\begin{tabular}{c  p{6.5cm} }
\hline
 {\bf Notation} & {\hspace{2.5cm}}{\bf Definition}
\\
\hline
$\Phi_{\mathrm{a}}$; $\lambda$ & PPP modeling the location of APs; the AP spatial deployment density \\
$K$; $N$; $G$ & Number of associated UEs per AP; number of subchannels; UE number over subchannel number ratio, i.e., $G=K/N$ \\
$P_{\mathrm{ut}}$; $\alpha$ & UE transmit power; path loss exponent \\
$\gamma_{k,t}$; $\theta$ & SINR received from UE $k$ at communication round $t$; the SINR decoding threshold \\
$\tilde{\rho}_{k,t}$; $\bar{\rho}_{k,t}$ & Instantaneous SNR of UE $k$ at communication round $t$; time average SNR of UE $k$ till communication round $t$ \\
$\mathcal{D}_k$; $n_k$ &  Data set of UE $k$; size of the data set $\mathcal{D}_k$ \\
$\ell_i(\cdot)$; $r(\cdot)$ &  Loss function on data point $\mathbf{x}_i$; regularization function \\
$\ell^*_i(\cdot)$; $r^*(\cdot)$ &  Conjugate function of $\ell_i(\cdot)$; conjugate function of $r^*(\cdot)$ \\
$\mu$; $\zeta$; $\kappa$ &  Smoothness of the loss function $\ell_i(\cdot)$; convexity of the regularizer $r(\cdot)$; partition difficulty of the data set \\
$P(\mathbf{w})$; $\mathbf{w}$ & The objective function; optimization vector of the primal problem \\
$D(\mathbf{a})$; $\mathbf{a}$ & The dual form of the objective function; the dual variables \\
$\eta^t$; $\beta$ & Local learning rate; error level of the local solution \\
$\mathcal{S}_{k,t}^z$; $\mathcal{U}^z_k$ & Indicator of the selection state of UE $k$ at communication round $t$, which takes value 1 if the UE is selected and 0 otherwise; parameter update success probability \\
\hline
\end{tabular}
\end{center}\vspace{-0.63cm}
\end{table}%
\section{System Model}\label{sec:sysmod}
In this section, we introduce the network topology and propagation model, the generic procedure of FL, and the scheduling policies.  The main notations used throughout the paper are summarized in Table~\ref{table:notation}.
\subsection{ Network Structure and Propagation Channel }
Let us consider a wireless network that consists of APs and UEs, as depicted in Fig.~\ref{fig:FL_WN}. The locations of APs follow a homogeneous PPP $\Phi_{ \mathrm{a} }$ with spatial density $\lambda$.
We assume each AP has $K$ associated UEs uniformly distributed within its Voronoi cell\footnote{ This is equivalent to the maximum average power association rule, and we fix the total number of UEs in each cell to simplify the notational complexity. Note that relaxing this assumption does not change the conclusions drawn from this paper.}.
In this network, a fixed amount of spectrum is equally divided into $N$ radio access channels, where $N < K$. We consider each AP is equipped with a single antenna and a computing processor.
For a generic UE $k$, we consider it is equipped with a single antenna and has a local data set $\mathcal{D}_k = \{ \mathbf{x}_i \in \mathbb{R}^d, y_i \in \mathbb{R} \}_{i=1}^{n_k}$ with $n_k = |\mathcal{D}_k|$ sample points, where $|\cdot|$ denotes the cardinality of a set.
Each UE also has the capability of performing local training.

In this network, all the UEs transmit with a constant power $P_{\mathrm{ut}}${\footnote{We unify the transmit power for notational simplicity. Nonetheless, note that the analysis of this paper can be extended to account for power control in a straightforward way \cite{ElSHos:14}.}}.
We adopt a block-fading propagation model, where the channels between any pair of antennas are assumed independent and identically distributed (i.i.d.) and quasi-static, i.e., the channel is constant during one transmission block and varies independently from block to block. We consider all propagation channels are narrow-band and affected by two attenuation components, namely the small-scale Rayleigh fading with unit mean power, and the large-scale path loss that follows a power law. Moreover, in consideration of spectral efficiency, we assume the whole spectrum is reused in every cell.

 \begin{table}
 \caption{ Loss functions for popular machine learning models} \label{table:LosFun}
 \begin{center}
 \renewcommand{\arraystretch}{1.3}
 \begin{tabular}{c  p{5.5cm} }
 \hline
  {\bf Model} & {\hspace{0.0cm}}{\bf Loss function $\ell_i( \mathbf{x}_i^T \mathbf{w} )$}
 \\
 \hline
 Smooth SVM & $\frac{1}{2} \max\{0, 1 - y_i \mathbf{w}^T \mathbf{x}_i \}$  \\
 Linear regression & $\frac{1}{2} \Vert y_i  - \mathbf{w}^T \mathbf{x}_i \Vert^2$  \\
 Logistic regression & $\log\big( 1 + \exp(  - y_i \mathbf{w}^T \mathbf{x}_i ) \big)$  \\
 K-means & $\frac{1}{2} \min_{ j \in \{ 1, 2, ..., K'\} } \Vert \mathbf{x}_i  - \mathbf{w}_j \Vert^2$, where $K'$ is the number of clusters \\
 Neural Network & $\frac{1}{2}\Vert y_i - \sum_{m=1}^M {v}_m \, \phi( \mathbf{w}_m^T \mathbf{x}_i ) \Vert$, where $\phi(\cdot)$ is the activation function, ${v}_m$ the weights connecting the neurons, and $M$ the number of neurons  \\
 \hline
 \end{tabular}
 \end{center}\vspace{-0.63cm}
 \end{table}%
\subsection{ Federated Learning }
At each AP, the goal is to learn a statistical model over data that reside on the $K$ associated UEs, i.e., the AP needs to fit a vector $\mathbf{w} \in \mathbb{R}^d$ so as to minimize a particular loss function by using the whole data set from all the UEs under its service. Formally, such task can be expressed as
\begin{align}\label{equ:Obj_Func}
\min_{ \mathbf{w} \in \mathbb{R}^d } \left\{ P( \mathbf{w} ) = \frac{1}{n} \sum_{ i=1 }^{n} \ell_i( \mathbf{x}^T_i \mathbf{w} ) + \xi r(\mathbf{w})  \right\}
\end{align}
where $n = \sum_{ i=1 }^K n_i$ is the size of the whole data set, $\xi$ is the regularizing parameter and $r(\mathbf{w})$ a deterministic penalty function. Common choices for $r(\mathbf{w})$ include the L-2 penalty $\Vert \mathbf{w} \Vert^2_2$, the L-1 penalty $\Vert \mathbf{w} \Vert_1$, or a family of folded concave functions \cite{Zha:10}.
The function $\ell_i(\cdot)$ represents the loss function associated with data point $\mathbf{x}_i$.
Several examples of loss functions used in popular machine learning models are summarized in Table~\ref{table:LosFun}.

If the data set $\mathcal{D} = \cup_{ k = 1 }^K \mathcal{D}_{k}$ is completely available at the AP, problem \eqref{equ:Obj_Func} can be easily solved via a number of machine learning algorithms.
However, such a data set is generally unavailable in a real-world setting because $i$) the amount of data at each UE can be large and the data uploading task may be constrained by energy and bandwidth limitations, and more importantly, $ii$) the data from to each UE may contain highly sensitive information, e.g., medical records, words typed in messager APPs, or web browsing history, and users are unwilling to share it.
As such, the FL algorithm has emerged, where the data collection process is decoupled from the global model training.
The general procedure of FL is summarized in Algorithm~1.
Particularly, each UE downloads a global model, $\mathbf{w}^t$, from the AP to conduct stochastic gradient descent (SGD) per equation (3), aiming to minimize the objective function $P(\mathbf{w})$ by only using information from the globally shared vector $\mathbf{w}^t$ and data set $\mathcal{D}_k$ (note that this data set is private).
The AP periodically collects all the trained parameters from UEs to produce a global average and then redistributes the improved model back to the UEs.
After a sufficient amount of training and update exchanges, usually termed communication rounds, between the AP and its associated UEs, the objective function \eqref{equ:Obj_Func} is able to converge to the global optimal.
When all the updates can be correctly received by the AP in every communication round, the convergence property of FL has been quantitatively demonstrated \cite{KonMcMRam:15}.
However, as the FL algorithm is generally run in a wireless setting where updates are sent through shared spectrum, which is unreliable due to random fading and inter-cell interference, updates from some UEs can be lost during the data transmission phase.
Moreover, the wireless medium is usually resource-constrained, and the AP thus needs to select a subgroup of UEs for parameter updates in each communication round.

Apart from the scheduling issue, the generic training approach per Algorithm~1 suffers potential setback of slow convergence \cite{ZhaFenYan:19WCM}, especially when the loss function or regularizer has a complicated form. Furthermore, the duration of local training in Algorithm~1 needs to be carefully designed so as to ensure the local solutions do not diverge from the global model \cite{WanTuoSal:19JSAC}. As a result, the local training period needs to be small and that may incur a large number of communication rounds which is not desirable.
In that respect, we propose an algorithm, which will be elaborated in Section III, that presents a more suitable alternative to train FL in a wireless setting.

\subsection{ Scheduling Policies }
In many real-world systems, communicating data between machines is several orders of magnitude slower than reading data from main memory and performing local computing \cite{LanLeeZho:17}. Hence, sequentially updating the trained parameters from all UEs before global aggregation as proposed in \cite{NisYon:18} can lead to large overhead in the communication time and is not desirable. Instead, the AP shall only select a subgroup of UEs and update their parameters simultaneously so as to keep the communication time within an acceptable range.
To this end, the scheduling policy plays a crucial role in assigning the resource-limited radio channels to the appropriate UEs.
In the following, we denote by $G = K/N$ the ratio of the number of UEs to the number of subchannels {\footnote{For simplicity, we assume $K$ is a multiple of $N$. In more general scenarios where $G=K/N$ is not an integer, we can choose $G = \lceil K/N \rceil$, where the $\lceil \cdot \rceil$ denotes the ceiling function.}} and consider three practical policies as our scheduling criteria \cite{YanWanQue:18,ChoBah:07}:
\begin{itemize}
  \item[(a)] \textit{Random Scheduling (RS)}: In each communication round, the AP uniformly selects the $N$ associated UEs at random for parameter update, each selected UE is assigned a dedicated subchannel to transmit the trained parameter.
  \item[(b)] \textit{Round Robin (RR)}: The AP arranges all the UEs into $G$ groups and consecutively assigns each group to access the radio channels and update their parameters per communication round.
  \item[(c)] \textit{Proportional Fair (PF)}: During each communication round, the AP selects $N$ out of the $K$ associated UEs according to the following policy:
  \begin{align}
  \mathbf{m}^* = \arg \!\!\!\!\!\!\!\!\!\max_{ \mathbf{m} \subset \{ 1, 2, ..., K \} } \left\{ \frac{ \tilde{\rho}_{m_1,t} }{ \bar{\rho}_{m_1,t} }, ..., \frac{ \tilde{\rho}_{m_N,t} }{ \bar{\rho}_{m_N,t} }  \right\}
  \end{align}
  where $\mathbf{m} = ( m_1, ..., m_N )$ is a length-$N$ vector and $\mathbf{m}^* = ( m_1^*, ..., m_N^* )$ represents the indices of the selected UEs, $\tilde{\rho}_{m_i,t}$ and $\bar{\rho}_{m_i,t}$ are the instantaneous and time average signal-to-noise ratio (SNR) of UE $m_i$ at the communication round $t$, respectively \cite{ChoBah:07}.
\end{itemize}

The following sections are devoted to the design of algorithms to run federated learning in wireless networks, as well as the analysis that quantifies the running time of FL under different scheduling policies.
\begin{algorithm}[t!]
\caption{ Generic Federated Learning Algorithm }
\begin{algorithmic}[1] \label{Alg:Gen_FL}
\STATE \textbf{Parameters:} $\tau$ = number of local steps per communication round, $\eta$ = step size for stochastic gradient descent.
\STATE \textbf{Initialize:} $\mathbf{w}^0 \in \mathbb{R}^d$
\FOR { $t = 0, 1, 2, ..., T-1$ }
\FOR { each UE $k \in \{ 1, 2, ..., K \}$ in parallel }
\STATE Initialize $\mathbf{w}_k^t = \mathbf{w}^t$
\FOR { $s$ = 1 to $\tau$ }
\STATE Sample $i \in \mathcal{D}_k$ uniformly at random, and update the local parameter $\mathbf{w}^t_k$ as follows
\begin{align} \label{equ:LocGraDscnt}
\mathbf{w}_k^t \!=\! \mathbf{w}_k^t \!-\! \eta ( \nabla \ell_i(\mathbf{w}_k^t) + \nabla r(\mathbf{w}^t) )
\end{align}
\ENDFOR
\STATE Send parameter $\mathbf{w}_k^t$ to the AP
\ENDFOR
\STATE The AP collects all the parameters $\{ \mathbf{w}^t_k \}_{k=1}^K$, and updates $\mathbf{w}^{t+1} = \frac{1}{n} \sum_{k=1}^K n_k \mathbf{w}^t_k $
\ENDFOR
\STATE \textbf{Output:} $\mathbf{w}^T$
\label{Alg1:Iteration_End}
\end{algorithmic}
\end{algorithm}
\section{ Distributed Computing and Parameter Update }\label{sec:Local_Comput}
In this section, we detail the procedure that decomposes the problem from \eqref{equ:Obj_Func} into a number of subproblems which can be solved by using only the local data at each UE. We also describe how the local training and update adapt to the scheduling policy. To facilitate the design and analysis, we make the following assumptions on the loss function and the regulator throughout this paper.
\begin{assumption}
\textit{
	The function $r: \mathbb{R}^d \rightarrow \mathbb{R}$ is $\zeta$-strongly convex, i.e., $\forall i \in \{ 1, ..., n \}$ and $\forall \, \mathbf{x}, \Delta \mathbf{x} \in \mathbb{R}^d$ it holds that
	\begin{align}
	r( \mathbf{x} + \Delta \mathbf{x} ) \geq r( \mathbf{x} ) + \nabla r( \mathbf{x} )^T \Delta \mathbf{x} + \frac{\zeta}{2} \Vert \Delta \mathbf{x} \Vert^2
	\end{align}
	where $\nabla r(\cdot)$ denotes the gradient of the function $r(\cdot)${\footnote{In this paper, we follow the convention and write the definition of strong convexity using the gradient \cite{Bub:15}. Nevertheless, note that strongly convex functions may not be differentiable, and in that case, one shall replace the gradient by the subgradient \cite{Roc:70}.}}.
}
\end{assumption}

\begin{assumption}
\textit{
  The functions $\ell_i: \mathbb{R} \rightarrow \mathbb{R}$ are $1/\mu$-smooth, i.e., $\forall i \in \{ 1, ..., n \}$ and $\forall \, x, \Delta x \in \mathbb{R}$ it holds that
  \begin{align}
  \ell_i( x + \Delta x ) \leq \ell_i(x) + \nabla \ell_i(x) \Delta x + \frac{1}{ 2 \mu } ( \Delta x )^2
  \end{align}
  where $\nabla \ell_i(\cdot)$ denotes the gradient of the function $\ell_i(\cdot)$.
}
\end{assumption}
\subsection{ Local Decomposition }
First of all, using the Fenchel-Rockafeller duality, we can express the local dual optimization problem of \eqref{equ:Obj_Func} in the following way.
\begin{lemma}
\textit{
  The optimization problem \eqref{equ:Obj_Func} can be rewritten in the following dual form:
  \begin{align} \label{equ:Dual_FedSGD}
  \max_{ \mathbf{a} \in \mathbb{R}^n } \left\{ D(\mathbf{a}) = - \sum_{ i=1 }^n \frac{ \ell^*_i( - a_i ) }{n} - \xi r^*( \frac{1}{\xi n} \mathbf{X} \mathbf{a} )  \right\}
  \end{align}
  where $\{ a_i \}_{i=1}^n \subset \mathbb{R}$ represents the set of the dual variables, $\mathbf{X} = [ \mathbf{x}_1, \mathbf{x}_2, ..., \mathbf{x}_n ] \in \mathbb{R}^{ d \times n }$ is the total data set, $\ell^*(\cdot)$ and $r^*(\cdot)$ are the convex conjugate functions of $\ell_i(\cdot)$ and $r(\cdot)$, respectively, given as follows:
  \begin{align}
  \ell_i^*( - a_i ) &= \sup_{ {u_i} \in \mathbb{R} } \{ - a_i u_i - \ell_i(u_i) \},\\
  r^*( \mathbf{a} ) &= \sup_{ \mathbf{s} \in \mathbb{R}^n } \{ \mathbf{s}^T \mathbf{a} - r(\mathbf{s}) \}.
  \end{align}
}
\end{lemma}
\begin{IEEEproof}
We first denote $\mathbf{u} = \mathbf{X}^T \mathbf{w}$. By using the Lagrangian, we can write the original problem \eqref{equ:Obj_Func} equivalently as follows:
\begin{align} \label{equ:Dual_Opt}
&\frac{1}{n} \min_{ \mathbf{u}, \mathbf{w} } \left\{ \sum_{ i=1 }^n \ell_i( \mathbf{x}_i^T \mathbf{w} ) + \xi n  r(\mathbf{w}) + \mathbf{a}^T ( \mathbf{u} - \mathbf{X}^T \mathbf{w} ) \right\}
\nonumber\\
= &\frac{1}{n} \inf_{ \mathbf{w} } \! \Big\{ \xi n r( \mathbf{w} ) - \mathbf{a}^T \mathbf{X}^T \! \mathbf{w} \Big\} +\!  \sum_{i=1}^n \inf_{ {u_i} } \! \Big\{ \ell_i(u_i) + a_i u_i  \Big\}
\nonumber\\
= & - \xi \sup_{ \mathbf{w} } \Big\{ \mathbf{w}^T \frac{ \mathbf{Xa} }{ \xi n } - r(\mathbf{w}) \Big\} -\! \sum_{i=1}^n \sup_{ {u_i} } \! \Big\{ \!\! - a_i u_i - \ell_i(u_i)   \Big\}
\nonumber\\
=& - \xi r^*\big( \frac{ 1 }{ \xi n } \mathbf{Xa} \big) - \sum_{i=1}^n \frac{ \ell_{i}^*(-a_i) }{ n } = D(\mathbf{a}).
\end{align}
Note that when $\mathbf{a}$ is chosen so as to maximize \eqref{equ:Dual_Opt}, the value of $D(\mathbf{a})$ is equivalent to \eqref{equ:Obj_Func} due to the first-order optimality condition \cite{Roc:70}. As such, the result in \eqref{equ:Dual_Opt} then follows from maximizing the above problem with respect to $\mathbf{a}$.
\end{IEEEproof}

The advantage of using the dual formulation in \eqref{equ:Dual_FedSGD} is that it allows us to better separate the global problem into a number of distributed subproblems solvable via federated computing across different UEs. In particular, we define $\mathbf{v}(\mathbf{a}) = \mathbf{Xa}/\xi n$ and first decompose $D(\mathbf{a})$ into the following form:
\begin{align} \label{equ:D_a}
D(\mathbf{a})
&=\! -\, \xi \, r^*( \mathbf{v}(\mathbf{a}) ) +  \sum_{k=1}^K \Big[  -\!\! \sum_{ i \in \mathcal{D}_k } \frac{ \ell_i^*( - a_i ) }{ n } \,\, \Big]
\nonumber\\
&= -\, \xi \, r^*( \mathbf{v}(\mathbf{a}) ) - \sum_{ k=1 }^K  R_{k}( \mathbf{a}_{[k]} )
\end{align}
where $R_{k}( \mathbf{a}_{[k]} ) = 1/n \sum_{ i \in \mathcal{D}_k } \ell^*_i( - {a}_i )$ with $\mathbf{a}_{ [k] } \in \mathbb{R}^n$ being the coordinates of the vector $\mathbf{a}$ that corresponds to the data set $\mathcal{D}_k$ and the other entries are set to zero.
As such, for a randomly initialized vector $\bar{\mathbf{a}}$, varying its value by $\Delta \mathbf{a}$ will result in the following change to \eqref{equ:D_a}:
\begin{align} \label{equ:D_delta_a}
D( \bar{\mathbf{a}} + \Delta \mathbf{a} ) =\! - \xi \, r^*( \mathbf{v}( \bar{\mathbf{a}} + \Delta \mathbf{a} ) ) -  \sum_{k=1}^K R_{k}( \bar{\mathbf{a}}_{[k]} + \Delta\mathbf{a}_{[k]}  ).
\end{align}
Notably, the changes in the second term of the above equation correspond to only the data set $\mathcal{D}_k$ of each local UE $k$, while the first term involves all the global variations. Because $r(\cdot)$ is $\zeta$-strongly convex, we know that $r^*(\cdot)$ is $1/\zeta$-smooth \cite[Theorem 4.2.1]{HirJeaLem:12} and can thus bound $r^*( \mathbf{v}( \bar{\mathbf{a}} + \Delta \mathbf{a} ) )$ as follows:
\begin{align} \label{equ:r_star_bnd}
r^*( \mathbf{v}( \bar{\mathbf{a}} + \Delta \mathbf{a} ) ) &\leq r^*( \frac{ \mathbf{X} \bar{\mathbf{a}} }{ \xi n } ) + \langle \frac{1}{ \xi n } \mathbf{X}^T \nabla r^*\!( \mathbf{v} ( \bar{\mathbf{a}} ) ) , \Delta \mathbf{a}  \rangle
\nonumber\\
&\quad + \frac{ \kappa }{ 2 ( \xi n )^2 } \big\Vert \mathbf{X} \Delta \mathbf{a} \big\Vert^2
\nonumber\\
&= r^*( \mathbf{v}( \bar{\mathbf{a}} ) ) + \sum_{ k=1 }^K \langle \frac{1}{ \xi n } \mathbf{X}_{[k]}^T \nabla r^*\!( \mathbf{v} ( \bar{\mathbf{a}} ) ) , \Delta \mathbf{a}_{[k]} \rangle \nonumber\\
&\quad + \frac{ \kappa }{ 2 ( \xi n )^2 } \sum_{k=1}^K \big\Vert \mathbf{X}_{[k]} \Delta \mathbf{a}_{[k]} \big\Vert^2,
\end{align}
where $\kappa > 1/\zeta$ is a data dependent term measuring the difficulty of the partition to the whole data set.
By substituting \eqref{equ:r_star_bnd} into \eqref{equ:D_delta_a} it yields
\begin{align} \label{equ:D_Delta_Ineq}
& D( \bar{\mathbf{a}} + \Delta \mathbf{a} ) \geq  - \xi r^*( \mathbf{v}( \bar{\mathbf{a}} ) ) - \sum_{ k=1 }^K \langle \frac{1}{ n } \mathbf{X}_{[k]}^T \nabla r^*\!( \mathbf{v} ( \bar{\mathbf{a}} ) ) , \Delta \mathbf{a}_{[k]} \rangle
\nonumber\\
& - \frac{ \kappa }{ 2 \xi n^2 } \sum_{k=1}^K \big\Vert \mathbf{X}_{[k]} \Delta \mathbf{a}_{[k]} \big\Vert^2 - \sum_{k=1}^K R_{k}( \bar{\mathbf{a}}_{[k]} + \Delta\mathbf{a}_{[k]}  ).
\end{align}
To this end, if each UE $k\in\{ 1, 2, ..., K \}$ can optimize $\Delta \mathbf{a}_{[k]}$ using its own data set $\mathcal{D}_k$ so as to maximize the right hand side (R.H.S.) of \eqref{equ:D_Delta_Ineq}, the resultant improvements can be combined to direct $D(\bar{\mathbf{a}})$ toward the optimal value{\footnote{Instead of directly solving the original optimization problem, we solve for an approximated surrogate which is advantageous due to the savings per communication round and the fact that solutions with extremely high accuracy are not necessary for machine learning in practice. }}.
To be more concrete, during any communication round $t$, the AP produces $\mathbf{v}( \mathbf{a}^t )$ by using updates received from the last round and broadcasts that to all the UEs.
The task at any given UE $k$ is to solve for $\Delta \mathbf{a}_{[k]}^t$ that maximizes the following:
\begin{align}\label{equ:Local_DelSuP}
&\Delta D_{k}( \Delta \mathbf{a}_{[k]}^t; \mathbf{v}( \mathbf{a}^t), \mathbf{a}_{ [k] }^t ) = - R_{k}( \mathbf{a}_{[k]}^t \!+\! \Delta \mathbf{a}_{ [k] }^t ) - \frac{ \xi }{ K } r^*\!( \mathbf{v}( \mathbf{a}^t) )
\nonumber\\
 & - \langle \frac{1}{n} \mathbf{X}_{[k]}^T \nabla r^*\!( \mathbf{v}(\mathbf{a}^t) ) , \Delta \mathbf{a}_{[k]}^t  \rangle - \frac{ \kappa / \xi }{ 2 n^2 } \, \big\Vert \mathbf{X}_{[k]} \Delta \mathbf{a}_{[k]}^t \big\Vert^2,
\end{align}
and then send the parameter $\Delta \mathbf{v}_k^t = \mathbf{X}_{[k]}^t \Delta \mathbf{a}_{[k]}^t / \xi n$ to the AP. The AP then updates the global vector as $\mathbf{v}(\mathbf{a}^t + \Delta \mathbf{a}^t) = \mathbf{v}(\mathbf{a}^t) + \sum_{k=1}^K \Delta \mathbf{v}_k^t$. As such, by alteratively updating $\mathbf{v}(\mathbf{a}^t)$ and $\{\Delta \mathbf{a}^t_{[k]}\}_{k=1}^K$ on the global and local sides, respectively, it is expected that the solutions to the dual problem can be enhanced at every step and that guarantees the original problem converges to the optimal.

It is important to note that unlike \eqref{equ:LocGraDscnt}, the subproblem \eqref{equ:Local_DelSuP} is simple in the sense that it is always a quadratic objective (apart from the $R_k(\cdot)$ term). The subproblem does not dependent on the function $r^*(\cdot)$ itself, but only its linearization at the shared vector $\mathbf{v}(\mathbf{a}^t)$. This property additionally simplifies the task of local solvers, especially when the function $r^*(\cdot)$ takes on a complicated form.
Moreover, if the local problems were solved exactly, this can be interpreted as a data-dependent block separable proximal step, which is known as a method to accelerate the learning process.

The requirement for such a decomposition method to work is that during each global aggregation, the changes in the local variables $\{\mathbf{a}^t_{[k]}\}_{k=1}^K$ on each UE and that in the global vector $\mathbf{v}(\mathbf{a}^t)$ are kept consistent [19].
However, because the wireless channels are generally unreliable, updates can be lost during the data transmission phase which leads to misalignment in the global and local parameters.
In the following, we will develop an algorithm that adapts the local training at each UE along with the communication condition in the global parameter updating phase.

\begin{figure*}[t!]
  \centering{}

    {\includegraphics[width=1.75\columnwidth]{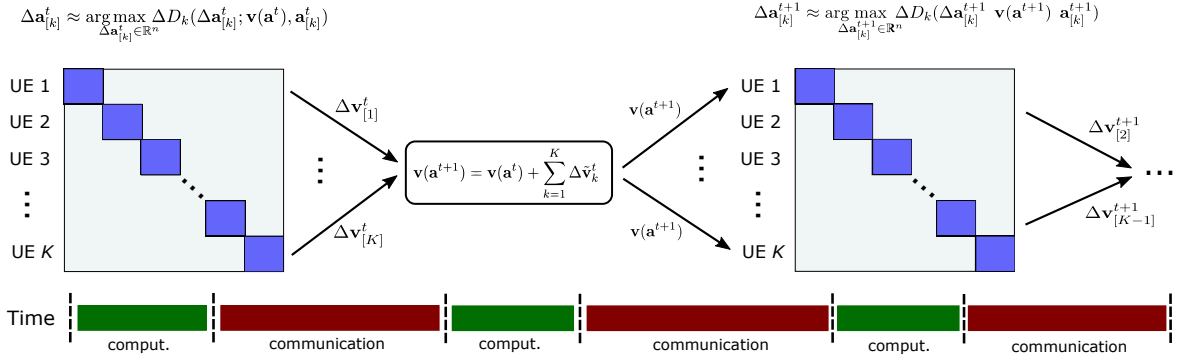}}

  \caption{ A typical iteration round of the learning procedure: $i$) each UE solves a subproblem using its locally stored data, $ii$) the AP selects a subgroup of UEs to collect updates based on which it produces an enhanced model, $iii$) the new model is sent back to the UEs, and the process is repeated. }
  \label{fig:Prc_FL}
\end{figure*}

\subsection{ Parameter Updates }
During a typical communication round $t$, in order to update the parameter $\Delta \mathbf{v}_k^t$ from a generic UE $k$ to the global AP, two conditions need to be simultaneously satisfied: $i$) the UE is selected by the AP, and $ii$) the transmitted data is successfully decoded.
In that respect, we first introduce $\mathcal{S}_{k,t}^z \in \{ 0, 1 \}$ as a selection indicator, with $z \in \{ \mathrm{RS}, \mathrm{RR}, \mathrm{PF} \}$ specifying the employed scheduling policy, where $\mathcal{S}_{k,t}^z = 1$ corresponds to the event that UE $k$ is chosen by the AP for transmission and $\mathcal{S}_{k,t}^z = 0$ otherwise.

Next, we characterize the transmission quality of the wireless links.
Note that although the depicted wireless network contains infinitely many APs, thanks to the stationary property of PPPs, the FL convergence rates of all APs are statistically equivalent.
As such, by applying Slivnyak's theorem to the stationary PPP of APs, it is sufficient to evaluate the SINR of
a typical AP at the origin \cite{BacBla:09,Hae:12}.
For signals transmitted from UE $k$ that is located at $c_k$, the SINR received at the typical AP takes the following form:
\begin{align}
\gamma_{k,t} = \frac{ P_{\mathrm{ut}} h_k \Vert c_k \Vert^{-\alpha} }{ \sum_{ c \in \tilde{\Phi}^k_{ \mathrm{u} } } P_{\mathrm{ut}} h_c \Vert c \Vert^{-\alpha} + \sigma^2  }
\end{align}
where $\alpha$ is the path loss exponent, $h_k \sim \exp(1)$ is the small scale fading, $\sigma^2$ is the variance of Gaussian additive noise, and $\tilde{\Phi}^k_{ \mathrm{u} }$ represents the locations of out of cell UEs that interfere with the typical AP.
In order for the AP to successfully decode the updates from UE $k$, it is required that the received SINR exceeds a decoding threshold $\theta$, i.e., $\gamma_{k,t} > \theta$.
Since the updated parameters from each UE have the same size, we assume the APs adopt a unified SINR decoding threshold in this network.

In any typical communication round, the probability of a generic UE being selected by its tagged AP depends on the scheduling policy employed. On the other hand, since both the signal strength and the interference received at a given AP are governed by a number of stochastic processes, e.g., the random spatial distribution of AP/UE locations and small-scale fading, the resulting SINR is a random variable.
As such, we define the following quantity, termed the \textit{parameter update success probability}, to characterize the transmission performance in each update
\begin{align} \label{equ:Upd_Succ}
\mathcal{U}^z_k = \mathbb{P}( \gamma_{k,t} > \theta, \mathcal{S}^z_{k,t} = 1 ), ~~ z \in \{ \mathrm{RS}, \mathrm{RR}, \mathrm{PF} \}.
\end{align}
This variable fully captures the key aspects for the successful update of parameters in each UE, and, as we will show later on, plays a critical role in the convergence analysis.

\begin{algorithm}[t!]
\caption{ Wireless Federated Learning Algorithm }
\begin{algorithmic}[1] \label{Alg:Wireless_FedSGD}
\STATE \textbf{Input:} Data set $\{ \mathcal{D}_k \}_{k=1}^K$ at the UEs, scheduling policy $z \in \{ \mathrm{RS}, \mathrm{RR}, \mathrm{PF} \}$ at the AP
\STATE \textbf{Initialization:} Each UE $k$ randomly initiates a starting point $\mathbf{a}_{[k]}^0 \in \mathbb{R}^n$. The AP randomly selects a portion of the associated UEs to collect $\mathbf{X}_{[k]}^T \mathbf{a}_{[k]}^0 / \xi n$, produces $\mathbf{v}(\mathbf{a}^0) := \mathbf{X}_{[k]}^T \mathbf{a}_{[k]}^0 / \xi n$, and sends the parameters $\mathbf{v}(\mathbf{a}^0)$ and $\eta^0 = K/2N$ to all the UEs
\FOR { $t = 0, 1, 2, ..., T-1$ }
\FOR { each UE $k \in \{ 1, 2, ..., K \}$ in parallel }
\STATE Compute $\mathbf{X}_{[k]}^T \nabla r^*(\mathbf{v}(\mathbf{a}^t)) $
\STATE Let $\Delta\mathbf{a}_{[k]}^t$ be an approximated solution of the local subproblem in \eqref{equ:Local_DelSuP}, i.e.,
\begin{align}
\Delta \mathbf{a}_{[k]}^t \approx \arg \!\!\!\!\! \max_{ \Delta \mathbf{a}_{[k]}^t \in \mathbb{R}^n } \!\!\!\! \Delta {D}_{k}( \Delta \mathbf{a}_{[k]}^t; \mathbf{v}(\mathbf{a}^t), \mathbf{a}_{ [k] }^t )
\end{align}
where $\kappa$ is chosen as $\kappa = K/\zeta$
\STATE Update and store the local reference parameter
\begin{align}\label{equ:ak_update}
\mathbf{a}_{ [k] }^{t+1} = \mathbf{a}_{ [k] }^{t} + \eta^t \Delta \mathbf{a}_{[k]}^t,
\end{align}
\STATE If $\mathcal{S}_{k,t}^z = 1$, compute the following global parameter and send it to AP via the allocated spectrum:
\begin{align}
\Delta \mathbf{v}_k^{t} = \frac{ 1 }{ \xi n } \mathbf{X}_{[k]} \Delta \mathbf{a}_{[k]}^t
\end{align}
otherwise, no update on the global parameter will be performed at the UE
\ENDFOR
\STATE The AP receives signals from the selected UEs, decodes the packets to extract each $\Delta \mathbf{v}_k^{t}$, and computes the improved parameter as
\begin{align} \label{equ:vt_update}
\mathbf{v}(\mathbf{a}^{t+1}) = \mathbf{v}(\mathbf{a}^t) + \sum_{k=1}^K \Delta \tilde{\mathbf{v}}_k^{t},
\end{align}
where $\Delta \tilde{\mathbf{v}}_k^{t}$ is given as
\begin{align}
\Delta \tilde{\mathbf{v}}_k^{t} = \left \{
  \begin{tabular}{cc}
  \!\!\!\!\! $\Delta {\mathbf{v}}_k^{t}$, & $\mathrm{if}~\mathcal{S}_{k,t}^z = 1 ~\mathrm{and}~ \gamma_{k,t} > \theta$,   \\
  \!\!\!\!\!\!  $0$, & \!\!\!\! \!\!\!\! \!\!\!\! \!\!\!\! $\mathrm{otherwise}$.
  \end{tabular}
  \right.
\end{align}
The AP also updates the variable $\eta_t$ as follows:
\begin{align} \label{equ:eta_t}
\eta^{t+1} = \frac{ t \times \eta^t }{t+1}  + \frac{ \sum_{k=1}^K \mathbbm{1}\{ \mathcal{S}_{k,t}^z = 1, \gamma_{k,t} > \theta \} }{ N(t+1) },
\end{align}
 and then broadcasts the updated global parameters $\mathbf{v}(\mathbf{a}^{t+1})$ and $\eta^{t+1}$ back to all the UEs.
\ENDFOR
\STATE \textbf{Output:} $\mathbf{ w }^T = \nabla r^*( \mathbf{v}(\mathbf{a}^T) )$.
\label{Alg1:Iteration_End}
\end{algorithmic}
\end{algorithm}
\subsection{ Federated Learning in Wireless Networks }
Armed with the above preparation, we are now ready to present the FL algorithm in a wireless network, which is summarized in Algorithm~\ref{Alg:Wireless_FedSGD} and illustrated by Fig.~\ref{fig:Prc_FL}.
We can see that the algorithm mainly consists of two parts:
\begin{itemize}
  \item At a typical UE $k$, it solves a local optimization problem \eqref{equ:Local_DelSuP} using only the data stored on the device. Based on the solution, the UE updates the local reference $\mathbf{a}_{[k]}^t$ per \eqref{equ:ak_update}, and if being selected by the AP, it sends out a global update $\Delta \mathbf{v}_{k}^t$ via the allocated subchannel.
  \item At the AP side, it selects a subgroup of UEs for update collection, decodes the received packet, and performs a global aggregation according to \eqref{equ:vt_update}. The new global parameter is redistributed to all the associated UEs using an error free channel.
\end{itemize}
Note that there is an incessant alternation between communication and computation during the training stage (cf. Fig.~2). In this regard, retransmissions of the failed packets may not be beneficial because each uplink transmission of local updates will be followed by a downlink transmission of the global average, and upon the reception of that, the UEs will refresh their reference parameters and start to solve a new subproblem using the local data{\footnote{If the transceivers are equipped with full duplex communications, it is possible to boost up the convergence rate because that has the potential to double the efficiency in both communication and computation aspects.}}.

Note that Algorithm~2 is essentially coordinate ascent working in the wireless setting. The crucial property here is that the optimization algorithm on UE $k$ changes only the coordinates of the dual optimization variable $\mathbf{a}_{[k]}^t$ corresponding to the data set $\mathcal{D}_k$.
Moreover, the factor $\eta^t$ acts as a time-averaging approach to calculate the parameter update success probability, which
steadily learns the quantity through the update status from each transmission.
As such, the update in \eqref{equ:ak_update} is able to adjust the local training along with the parameter update quality.
To be more concrete, under good channel conditions, the updates from UEs can be successfully received in each communication round, which leads to high value of the quantity $\eta^t$, indicating that the local references $\{ \mathbf{a}^t_{[k]} \}_{k=1}^K$ can progress more aggressively.
On the contrary, when the UEs are under a disadvantageous communication environment, the local learning rate $\eta^t$ also declines automatically, making the progress of local training more conservative.
This is because when communications are not reliable, the AP normally only receives a few updates from the UEs, which results in small changes in the global vector $\mathbf{v}(\mathbf{a}^t)$. In correspondence, local references shall not change abruptly but rather maintain the changes in line with the global ones{\footnote{Note that it is possible to prove the convergence of Algorithm~2 when $\eta^t$ is set differently. Nevertheless, the value of this quantity affects the ultimate rate of convergence \cite{JagSmiJor:15}.}}.

\remark{ \textit{The main benefit of Algorithm 2 arises from three properties: $i$) it is based on local second-order information and does not require sending gradients and Hessian matrices to the AP, which would be a significant cost in terms of communication, $ii$) the local subproblems are in the form of proximal steps, which can potentially accelerate the convergence rate, and $iii$) the local step size adjusts in accordance with the communication environment.}
}

\remark{ \textit{ While methods in \cite{MaKonJag:17} have a similar structure to Algorithm~\ref{Alg:Wireless_FedSGD}, they require the changes in local variables $\mathbf{a}_{ [k] }^{t}$ from each UE and the global change in $\mathbf{v}^t$ to be kept consistent, i.e., $\mathbf{v}^t = 1 / \xi n \mathbf{X} \mathbf{a}_{ [k] }^{t}$, which may hardly be satisfied in situations where communication is unreliable.
In contrast, Algorithm~\ref{Alg:Wireless_FedSGD} allows local updates to be asynchronized with the global aggregation. In fact, as we will show in Section~IV, as long as the local and global updates are aligned in an average manner, the FL is guaranteed to converge.
  }
}

\remark{\textit{In certain scenarios, e.g., the AP is training a support vector machine (SVM) with the UEs under ideal communication conditions, namely $N=K$ and $\theta=0$. The advantage of Algorithm~2 over Algorithm~1 is clear due to $a$) the subproblem (14) exactly matches the dual format and $b$) the partial solutions can be attained by means of second-order methods, which has a competitive edge of achieving faster convergence rate, rather than the SGD.}
}
\section{ Convergence Analysis }\label{sec:Convg_Ana}
In the following, we provide a quantitative analysis of the convergence properties of our proposed algorithm under various scheduling schemes. We also investigate two special cases to develop further insights. For better readability, most proofs and mathematical derivations have been relegated to the appendix.
\subsection{Preliminaries}
First of all, by using the first-order optimality condition, a mapping between the dual variable $\mathbf{a} \in \mathbb{R}^n$ and the primal candidate vector $\mathbf{w} \in \mathbb{R}^d$ exists and can be expressed as follows:
\begin{align}
\mathbf{w}(\mathbf{a}) = \nabla r^*( \mathbf{v}(\mathbf{a}) ) = \nabla r^*( \mathbf{X} \mathbf{a} /n ).
\end{align}
From strong duality we know that if $\mathbf{a}^*$ is an optimal solution of \eqref{equ:Dual_FedSGD}, then $\mathbf{w}(\mathbf{a}^*)$ is an optimal solution of \eqref{equ:Obj_Func},  i.e., the following duality gap holds:
\begin{align}
P( \mathbf{w}(\mathbf{a}^*)) - D(\mathbf{a}^*) = 0,
\end{align}
which ensures that by solving the dual problem \eqref{equ:Dual_FedSGD} we also solve the original primal problem of interest \eqref{equ:Obj_Func}. To this end, it is sufficient to use the gap between primal-dual as a measure of solution quality.

Next, note that $\eta^t$ in \eqref{equ:eta_t} can be rewritten as follows:
\begin{align}
\eta^t = \frac{1}{ N t } \sum_{ l=0 }^{t-1} \sum_{ k=1 }^K \mathbbm{1}\{ \mathcal{S}_{k,l}^z = 1, \gamma_{k,l} > \theta \}.
\end{align}
By noticing that the updates are i.i.d. and using the law of large numbers, we arrive at the following relationship:
\begin{align}
\mathcal{U}^z_{k} = \lim_{ t \rightarrow \infty } \frac{1}{ N t } \sum_{ l=0 }^{t-1} \sum_{ k=1 }^K \mathbbm{1}\{ \mathcal{S}_{k,l}^z = 1, \gamma_{k,l} > \theta \},
\end{align}
which is equivalent to that $\mathcal{U}_{k}^z = \mathbb{E}[ \eta^t ]$.

As such, we are able to evaluate the expected change in the dual objective function in \eqref{equ:Dual_FedSGD} over any typical communication round.
\begin{lemma}\label{lma:Convg_Bnd}
\textit{
  At any iteration $t$, with parameters $\mathbf{a}_{[k]}^{t+1}$, $\Delta \mathbf{a}_{k}^t$, and $\Delta \mathbf{v}_{k}(\mathbf{a}_{[k]}^t)$, $k\in\{1,2,...,K\}$, being updated according to Algorithm~\ref{Alg:Wireless_FedSGD}, the following condition holds:
  \begin{align} \label{equ:ExpctImprv}
 &\mathbb{E}\big[ D( \mathbf{a}^{t+1} ) \big] \!\geq\!  \sum_{k=1}^K \Big[\, \mathcal{U}^{ z }_k \Delta D_k( \Delta \mathbf{a}_{[k]}^t; \mathbf{v}(\mathbf{a}^t), \mathbf{a}^t_{[k]} )
 \nonumber\\
 & \quad   + \big( 1 - \mathcal{U}^{ z }_k \big) D( \mathbf{a}^t )/K \Big], \quad \forall z \in \{ \mathrm{RS}, \mathrm{RR}, \mathrm{PF} \}.
  \end{align}
}
\end{lemma}
\begin{IEEEproof}
See Appendix~\ref{apx:Convg_Bnd}.
\end{IEEEproof}
This result lies at the core of our convergence analysis because it allows us to quantify the impact of different scheduling policies on the updates of the objective function. It can be observed from \eqref{equ:ExpctImprv} that for a scheduling scheme that provides higher parameter update success probability, there is also larger potential to improve the objective function, and vice versa.

On the other side, since the trained parameters are periodically collected by the AP, UEs will need to finish their local computing before a given deadline. Due to the heterogeneity in the local computing environment, e.g., the difference in the size of the data sets or the computational capabilities, some UEs may not be able to obtain the optimal local solution upon the time for global updating. As such, we introduce the error level and make the following assumption.
\begin{assumption}\label{appm:prec_levl}
\textit{
  During each iteration $t$, we assume all the UEs can solve their local problem with error level $\beta \in ( 0, 1 )$, i.e., $\forall k \in \{1, 2, ..., K\}$, the following holds:
  \begin{align}
  \Delta D_k( \Delta \mathbf{a}^*_{[k]}; \mathbf{v}(\mathbf{a}^t), \mathbf{a}^t_{[k]} ) - \Delta D_k( \Delta \mathbf{a}^t_{[k]}; \mathbf{v}(\mathbf{a}^t), \mathbf{a}^t_{[k]} )
\nonumber\\
  \leq \beta \Big[  \Delta D_k( \Delta \mathbf{a}^*_{[k]}; \mathbf{v}(\mathbf{a}^t), \mathbf{a}^t_{[k]} ) - \Delta D_k(  \mathbf{0}; \mathbf{v}(\mathbf{a}^t), \mathbf{a}^t_{[k]} ) \Big]
  \end{align}
  where $\Delta \mathbf{a}^*_{[k]}$ is the minimizer of subproblem \eqref{equ:Local_DelSuP}.
}
\end{assumption}

The error level represents the quality of local computing, whereas in the above assumption we limit the quality of all the local solutions to be within a certain range. Note that the value of the error level $\beta$ can actually change across time, while we fix it as a constant for the sake of facilitating the analysis.
With all these results on hand, we are able to investigate the effect of scheduling methods on federated learning.
\subsection{Analysis}
We now analyze the convergence of FL operating in wireless systems.
In particular, we quantify the convergence rate of an FL algorithm using the number of required communication rounds such that the primal and dual problems can reach a certain duality gap, since upon that the trained parameter can be guaranteed to be in the vicinity of the optimal solution.
This brings us to the main theoretical result of this paper.
\begin{theorem} \label{thm:CnvRate_Genl}
\textit{
For any convergence target $\varepsilon$, the FL running under Algorithm~2 is able to achieve an $\varepsilon$ duality gap after $T_z$ rounds of communication, i.e.,
\begin{align}
\mathbb{E}[ P(\mathbf{w}(\mathbf{a}^{T_{ z } } ) ) - D(\mathbf{a}^{ T_{ z } }) ] < \varepsilon
\end{align}
if $T_z$ satisfies the following
\begin{align} \label{equ:TzBound}
T_z \geq \frac{ \log ( \varepsilon / n ) }{ \log \big( 1 - (1-\beta) \,\mathcal{U}^z_k \big)  }, ~~~z\in \{ \mathrm{RS}, \mathrm{RR}, \mathrm{PF} \}.
\end{align}
}
\end{theorem}
\begin{IEEEproof}
See Appendix~\ref{apx:CnvRate_Genl}.
\end{IEEEproof}
The above theorem demonstrates the general convergence property of FL in wireless networks. Using \eqref{equ:TzBound}, we can summarize the roles of iteration algorithms and scheduling policies in the remark below.
\remark{\textit{Due to the gradient descent (GD) based training approach, iteration complexities under all the scheduling policies are on the same order of GD's complexity, i.e., $\log(n/\varepsilon)$, while different scheduling policies affect the multiplicity constant, i.e., $\mathcal{U}^z_k$.}
}

Based on Theorem~\ref{thm:CnvRate_Genl}, we analyze and compare the convergence rate of FL running under three different scheduling policies, i.e., RS, RR, and PF, in the following.
\subsubsection{ Random Scheduling Policy } Selecting UEs uniformly at random for the update is the simplest and most widely adopted approach in practice. This method does not leverage any information from either the computing stage or the channel state. The following result characterizes the FL convergence performance under this method.
\begin{corollary} \label{Cor:Cnv_RS}
\textit{Under the RS policy, the parameter update success probability from a typical UE is given by
  \begin{align}
  \mathcal{U}^{ \mathrm{RS} }_k \approx \frac{ 1/G }{ 1 +  \mathcal{V}(\theta, \alpha)  }
  \end{align}
  where $\mathcal{V}(\theta, \alpha)$ is given as
  \begin{align} \label{equ:V_alpha}
   \mathcal{V}(\theta, \alpha) = \frac{ \sigma^2 \theta \lambda^{1 - \frac{\alpha}{2}} }{ \PVT 2^{\alpha-2} } + \theta^{ \frac{ 2 }{ \alpha } } \!\!\! \int_0^\infty \! \frac{ 1 - e^{ - \frac{12}{ 5 \pi } \theta^{ \frac{2}{\alpha} } u} }{ 1 + u^{ \frac{\alpha}{2} } } du.
  \end{align}
  Hence, by choosing the $T_{ \mathrm{RS} }$ such that
  \begin{align} \label{equ:RS_Rate}
  T_{ \mathrm{RS} } & \geq \frac{ \log( { \varepsilon }/{ n } ) }{ \log\Big( { 1 - \frac{ (1-\beta)/G }{ 1 + \mathcal{V}(\theta, \alpha) } } \Big) },
  \end{align}
  the expected duality gap satisfies
  \begin{align}
  \mathbb{E}[ P(\mathbf{w}(\mathbf{a}^{T_{ \mathrm{RS} } } ) ) - D(\mathbf{a}^{ T_{ \mathrm{RS} } }) ] < \varepsilon.
  \end{align}
}
\end{corollary}
\begin{IEEEproof}
See Appendix~\ref{apx:Cnv_RS}.
\end{IEEEproof}

It is noteworthy that the term $\mathcal{V}(\theta, \alpha)$ can be intuitively interpreted as the average interference plus noise power over the weighted received signal power, where the weight is proportional to $1/\theta$.
As such, $\mathcal{V}(\theta, \alpha)$ can be regarded as a metric to gauge the difficulty of decoding. In particular, when $\theta$ is small, the power of the desired signal is amplified and that gives a higher chance for the AP to successfully decode the signal.
This results in a small value of $\mathcal{V}(\theta, \alpha)$ and vice versa. Analogously, when $\alpha$ is small, that gives rise to higher interference levels which deteriorates the decoding process. And this fact is also reflected in an increase of $\mathcal{V}(\theta, \alpha)$.

\subsubsection{Round Robin Policy}
Unlike RS, the RR is operated under strict control and provides short-term fairness for all the UEs, i.e., each UE is guaranteed to update its parameter in a sequential way. This fairness property is captured in the following corollary.
\begin{corollary} \label{Cor:Cnv_RR}
\textit{Under the RR policy, the parameter update success probability from a typical UE is given by
  \begin{align}\label{equ:ActProb_Gnrl}
  \mathcal{U}^{ \mathrm{RR} }_k \approx \left \{
  \begin{tabular}{cc}
  \!\! $\frac{ 1 }{ 1  + \mathcal{V}(\theta, \alpha)  } $, & $\mathrm{if~scheduled}$,   \\
  \!\!\!\!  $0$, & $\mathrm{otherwise}$
  \end{tabular}
  \right.
  \end{align}
  where $\mathcal{V}(\theta, \alpha)$ is given in \eqref{equ:V_alpha}.
  Hence, by choosing the $T_{ \mathrm{RR} }$ such that
  \begin{align}
  T_{ \mathrm{RR} } & \geq \frac{ G \, { \log( \varepsilon / n ) } }{ \log\Big( { 1 - \frac{ 1 - \beta }{ 1 + \mathcal{V}(\theta, \alpha) } } \Big) },
  \end{align}
  the expected duality gap satisfies
  \begin{align}
  \mathbb{E}[ P(\mathbf{w}(\mathbf{a}^{ T_{ \mathrm{RR} } }) ) - D(\mathbf{a}^{ T_{ \mathrm{RR} } }) ] < \varepsilon.
  \end{align}
}
\end{corollary}
\begin{IEEEproof}
See Appendix~\ref{apx:Cnv_RR}.
\end{IEEEproof}

\subsubsection{Proportional Fair Policy}
When using PF as a scheduling policy, the AP can leverage additional information from the channel state for the UE selection.
Intuitively, there will be an improvement in the parameter update probability via PF, and the following result confirms such intuition.
\begin{corollary} \label{Cor:Cnv_PF}
\textit{Under the PF policy, the parameter update success probability from a typical UE is given by
  \begin{align}
  \mathcal{U}^{ \mathrm{PF} }_k \approx \!\! \sum_{ i=1 }^{ K \!-\! N \!+\! 1 } \! \binom{ K \!-\! N \!+\! 1 }{ i } \, \frac{ (-1)^{i+1} / G }{ 1 + \mathcal{V}( i\theta, \alpha ) }
  \end{align}
  where $\mathcal{V}(\theta, \alpha)$ is given in \eqref{equ:V_alpha}.
  Hence, by choosing the $T_{ \mathrm{PF} }$ such that
  \begin{align}
  T_{ \mathrm{PF} } & \geq
  \frac{ \log( \varepsilon / n ) }{ \log\!\Big( 1 \!-\! ( \,1 \!-\! \beta \, ) \sum_{ i=1 }^{ K \!-\! N \!+\! 1 } \! \binom{  K \!-\! N \!+\! 1 }{ i } \, \frac{ (-1)^{i+1} / G }{ 1 + \mathcal{V}( i\theta, \alpha ) } \Big) },
  \end{align}
  the expected duality gap satisfies
  \begin{align}
  \mathbb{E}[ P(\mathbf{w}(\mathbf{a}^{ T_{ \mathrm{PF} } }) ) - D(\mathbf{a}^{ T_{ \mathrm{PF} } }) ] < \varepsilon.
  \end{align}
}
\end{corollary}
\begin{IEEEproof}
See Appendix~\ref{apx:Cnv_PF}.
\end{IEEEproof}

Several remarks regarding Corollaries 1 to 3 are in order.
\remark{\textit{The convergence rate of FL degrades monotonically with an increase in the number of UEs per AP, $K$, since the additional UEs exacerbate the competition for communication resources and that deteriorates the parameter update probability of each UE. }}

\remark{\textit{When the wireless system is operating under high SINR threshold, i.e., $\theta \gg 0~$dB, in order to achieve an $\varepsilon$ duality gap, the required communication rounds of FL running under RS, RR, and PF are respectively given as follows:
  \begin{align}
  T_{ \mathrm{RS} } &\gtrsim  G \log( n/\varepsilon )  \frac{ 1  + \mathcal{V}(\theta, \alpha)  }{ 1 - \beta }, \\ \label{equ:T_RR_high_SINR}
  T_{ \mathrm{RR} } &\gtrsim  G \log( n/\varepsilon )  \frac{ 1  + \mathcal{V}(\theta, \alpha)  }{ 1 - \beta }, \\
  T_{ \mathrm{PF} } &\gtrsim  \frac{ \log( n/\varepsilon ) }{ N(1-1/G) + 1/G }  \frac{ 1  + \mathcal{V}(\theta, \alpha)  }{ 1 - \beta }.
  \end{align}
  It can be seen that in the high SINR regime, the RS and RR policies have similar convergence performance, while the PF policy converges more rapidly.
  }
} \label{rmk:High_SINR}

\remark{\textit{When the wireless system is operating under low SINR threshold, i.e., $\theta \ll 0~$dB, in order to achieve an $\varepsilon$ duality gap, the required communication rounds of FL running under RS, RR, and PF are respectively given as follows:
  \begin{align}
  T_{ \mathrm{RS} } &\gtrsim \frac{ \log( \varepsilon / n )  }{ \log( 1 - \frac{1-\beta}{G} ) }, \\
  T_{ \mathrm{RR} } &\gtrsim \frac{  G \log(  \varepsilon / n ) }{ \log( \beta ) }, \\
  T_{ \mathrm{PF} } &\gtrsim \frac{ \log( \varepsilon / n )  }{ \log( 1 - \frac{1-\beta}{G} ) }.
  \end{align}
  It can be seen that in the low SINR regime, the RS and PF policies have similar convergence performance, while the RR policy converges more rapidly.
  }
}\label{rmk:Low_SINR}

\subsection{Special Cases}
By leveraging the mathematical framework above, we are able to further consider three special cases: $a$) one shot communication, i.e., the UEs update their parameters in a one-by-one sequential order (which is equivalent to taking $G=K$ in the RR policy) and the APs allocate all the spectrum for the transmission in each communication round, $b$) multi-round communication, in which the AP waits several communication rounds to collect more updates before one global aggregation is performed, and $c$) all at once communication, namely all the UEs simultaneously access the spectrum during each communication round without any scheduling policy being employed.

We first characterize the convergence of FL under one shot communication.
\begin{corollary}
\textit{
  When parameters from each UE are updated via one shot communication, for a given convergence target $\varepsilon$, by choosing the training time $T_{ \mathrm{OS} }$ such that
  \begin{align} \label{equ:OS_Rate}
  T_{ \mathrm{OS} } & \geq
  \frac{ K \log( \varepsilon / n ) }{ \log\!\Big( 1 \!-\! \frac{ 1 - \beta }{ 1 + \mathcal{V}(\theta/N, \alpha) } \Big) },
  \end{align}
  the expected duality gap satisfies
  \begin{align}
  \mathbb{E}[ P(\mathbf{w}(\mathbf{a}^{ T_{ \mathrm{NS} } }) ) - D(\mathbf{a}^{ T_{ \mathrm{NS} } }) ] < \varepsilon.
  \end{align}
}
\end{corollary}
\begin{IEEEproof}
This result easily follows by noticing that under one shot communication, UEs can access the whole spectrum once in every $K$ communication rounds, and since every UE fully utilize the spectrum for its transmission, the required SINR threshold reduces to $(1+\theta)^{1/N} - 1 \approx \theta/N$.
\end{IEEEproof}

This corollary delivers a twofold message: $i$) the FL can perform very robustly in wireless system, where even the updates from each UE are sent far apart in time (proportional to the total UE number), the scheme is still guaranteed to converge, and $ii$) packing more UEs into each communication round facilitates faster convergence, which can be observed by comparing \eqref{equ:RS_Rate} and \eqref{equ:OS_Rate} and notice that even RS can largely outperform one shot communication in terms of the convergence rate. Hence, being able to collect updates from more UEs is more desirable than getting a small number of updates but in a highly reliable manner, which confirms the intuition and empirical approaches of packing more UEs into the spectrum during each communication round \cite{YanJiaShi:18,NisYon:18}.

Next, we study the effect of multi-round communication on the FL convergence rate. To be formal, let us denote by $C$ a divisor of $K$ and assume the AP adopts the RS as its scheduling policy. With $C$ rounds of update transmissions before each global aggregation, the FL has the following convergence performance.
\begin{corollary}
\textit{
  When parameters are updated under multi-round communication, for any given convergence target $\varepsilon$, by choosing the $T_{\mathrm{MC}}$ such that
  \begin{align}
    T_{\mathrm{MC}} \geq \frac{ C \log( \varepsilon /n ) }{ \log \big( 1 - \frac{ (1-\beta) C / G }{ 1 + \mathcal{V}(\theta, \alpha) } \big) },
  \end{align}
  the expected duality gap satisfies
  \begin{align}
  \mathbb{E}[ P(\mathbf{w}(\mathbf{a}^{ T_{ \mathrm{MC} } }) ) - D(\mathbf{a}^{ T_{ \mathrm{MC} } }) ] < \varepsilon.
  \end{align}
}
\end{corollary}
\begin{IEEEproof}
Note that under such a scheme, both the parameter update success probability and the required communication rounds are increased by a factor of $C$. The result then follows by leveraging an approach similar to the proof of Corollary~\ref{Cor:Cnv_RS}.
\end{IEEEproof}

The equation above reveals that the gain from enhanced communication reliability cannot compensate for the loss of degrees of freedom in the time domain. Therefore, waiting for more updates before the global aggregation is not desirable if that incurs additional communication rounds.
This result also provides theoretical support to our claim in Section II-C that sequentially updating parameters from all the UEs before global aggregation is not desirable in FL.

Finally, when no schedule is asserted, i.e., all the UEs can access the spectrum simultaneously during each communication round, it increases the efficiency of channel use for each UE while also giving rise to a higher level of mutual interference. To simplify the notational complexity, we assume each subchannel has $G$ simultaneously transmitting UEs. The following corollary then describes the convergence in such a scenario.
\begin{corollary}
\textit{
When parameters are updated via all at once communication, for any given convergence target $\varepsilon$, by choosing the $T_{ \mathrm{NS} }$ such that
  \begin{align}
  T_{ \mathrm{NS} } & \geq
  \frac{ \log( \varepsilon / n ) }{ \log\!\Big( 1 \!-\! \frac{ 1 - \beta }{ 1 + \mathcal{Z}(\theta, \alpha) } \Big) },
  \end{align}
  where $\mathcal{Z}(\theta, \alpha)$ is given as
  \begin{align}
    \mathcal{Z}(\theta, \alpha) = \frac{ \theta \sigma^2 \lambda^{ \frac{\alpha}{2} } }{ \PVT 2^{ \frac{\alpha}{2} - 1 } } + G \!\! \int_0^\infty \!\! \frac{ \theta^{ \frac{2}{\alpha} } du}{ 1 + u^{\alpha/2} },
  \end{align}
  the expected duality gap satisfies
  \begin{align}
  \mathbb{E}[ P(\mathbf{w}(\mathbf{a}^{ T_{ \mathrm{NS} } }) ) - D(\mathbf{a}^{ T_{ \mathrm{NS} } }) ] < \varepsilon.
  \end{align}
}
\end{corollary}
\begin{IEEEproof}
When no scheduling is asserted, we have $\mathbb{P}(\mathcal{S}_{k,t}=1)=1$ and the SINR received at UE $k$ can be written as
\begin{align}
\gamma_{k,t}^{ \mathrm{NS} } = \frac{ \PVT h_k \Vert c_k \Vert^{-\alpha} }{ \sum_{ c \in {\Phi}_{ \mathrm{u} }^{ \mathrm{NS} } } \PVT h_c \Vert c \Vert^{-\alpha} + \sigma^2  },
\end{align}
where ${\Phi}_{ \mathrm{u} }^{ \mathrm{NS} }$ is the set of locations of interfering UEs under the all at once communication. By Slivnyark's theorem \cite{BacBla:09}, the interfering points form a PPP with spatial density $\lambda G$ and the transmission success probability can be calculated as
\begin{align} \label{equ:SucProb_Cond_NS}
& \mathbb{P}( \gamma_{k,t}^{ \mathrm{NS} } > \theta | r_k, \mathcal{S}_{k,t} = 1 )
\nonumber\\
&= \mathbb{E}\Big[ \exp\!\big(\! -\! \frac{ \theta \sigma^2 r_k^\alpha }{ \PVT }  - \!\! \int_0^\infty\!\!\!\! \frac{ \lambda G \pi }{ 1 \!+\! \Vert x \Vert^\alpha / \theta r_k^\alpha } dx \Big) \Big]
\nonumber\\
& \approx \exp\!\Big(  \frac{ - \theta \sigma^2 r_k^2 }{ \PVT (2 \lambda)^{ \frac{ \alpha }{ 2 } - 1 } } -\! \lambda G \pi r_k^2 \theta^{ \frac{ 2 }{ \alpha } } \!\!\! \int_0^\infty \! \frac{ 1 }{ 1 + u^{ \frac{\alpha}{2} } } du \Big).
\end{align}
The result follows by deconditioning \eqref{equ:SucProb_Cond_NS} with respect to \eqref{equ:Dist_R0} to obtain the parameter update success probability, and then using a similar approach per Corollary~\ref{Cor:Cnv_RS} to show the necessary iterations for a desired duality gap.
\end{IEEEproof}

Note that the quantity $\mathcal{Z}(\theta, \alpha)$ plays a similar role as $\mathcal{V}(\theta, \alpha)$, with the interference counted from different regions.
We further note that when the network is operating under very low SINR threshold, i.e., $\theta \ll 0$ dB, then $\mathcal{Z}(\theta, \alpha)\approx \mathcal{V}(\theta, \alpha)$ and the required iterations to achieve duality gap $\varepsilon$ is
  \begin{align}\label{equ:T_NS_high_SINR}
  T_{ \mathrm{NS} } &\gtrsim   \frac{ \log( n/\varepsilon )  }{ \log(\beta) }.
  \end{align}
By comparing \eqref{equ:T_RR_high_SINR} with \eqref{equ:T_NS_high_SINR}, we can observe that in the very low SINR regime, round robin scheduling performs not even as good as naively transmitting the parameters from all UEs simultaneously, i.e., no schedule, showing the importance of choosing appropriate scheduling methods in different regimes.

Several numerical results based on the analysis derived in this section will be shown in Section V to give more practical insights into the design of scheduling schemes for federated learning in wireless networks.
\begin{figure*}[t!]
  \centering

  \subfigure[\label{fig:1a}]{\includegraphics[width=0.95\columnwidth]{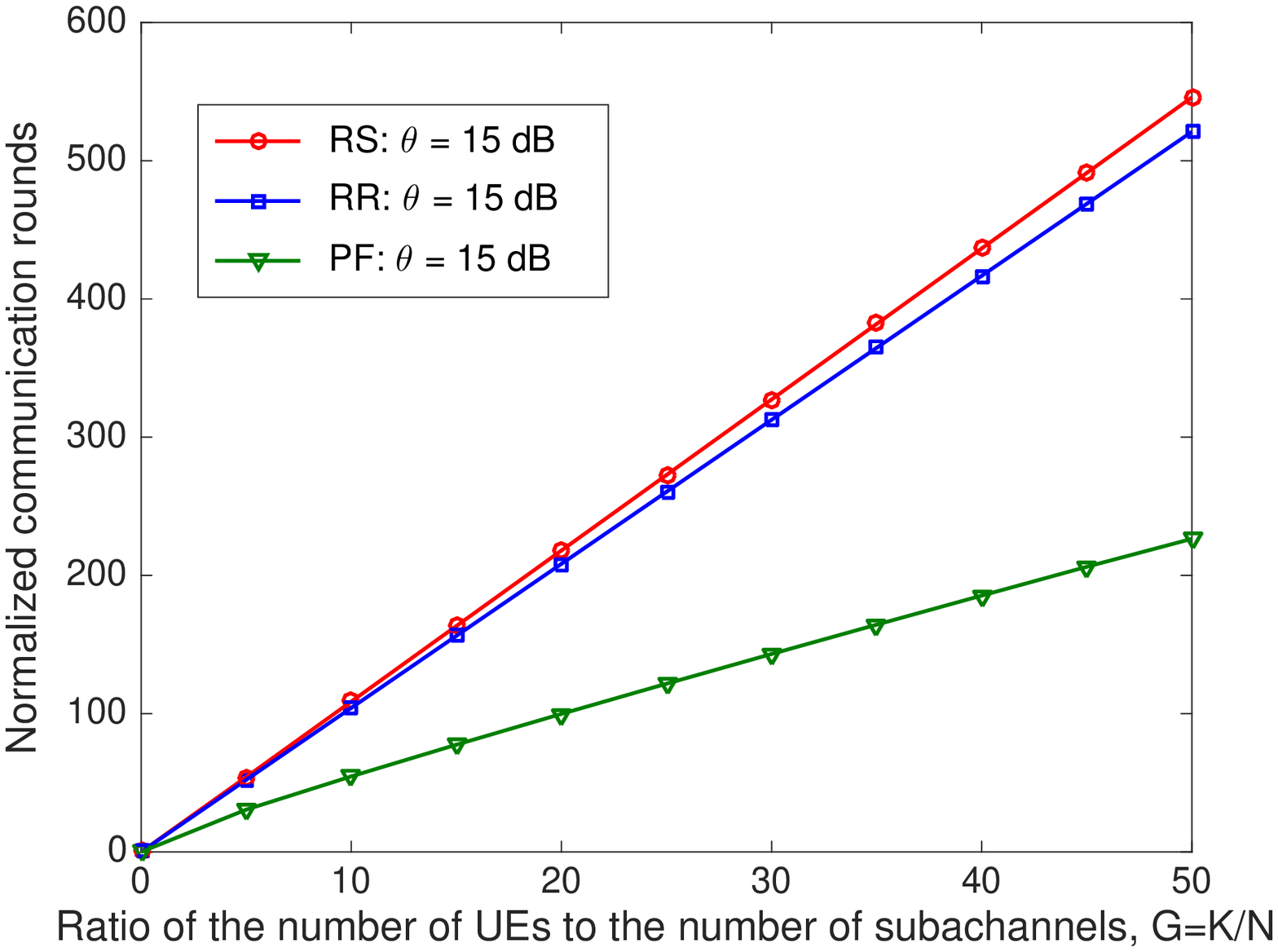}}~\qquad\quad
  \subfigure[\label{fig:1b}]{\includegraphics[width=0.95\columnwidth]{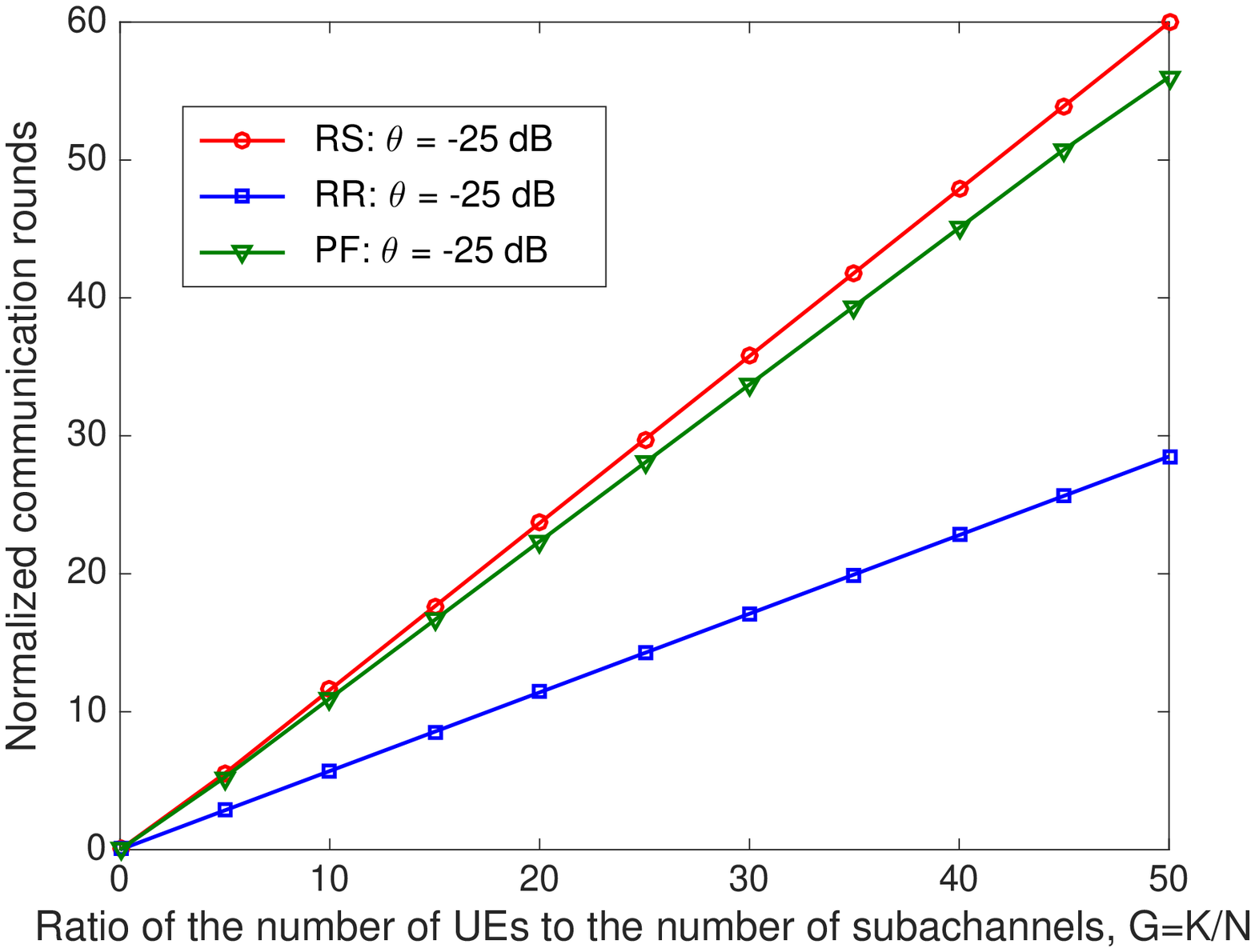}}
  \caption{ Normalized communication rounds vs UE number over subchannel number ratio, the number of subchannels is set as $N=10$. In Fig. (a), we plot the normalized communication rounds under high SINR threshold regime. In Fig. (b), we depict the normalized communication rounds under low SINR threshold regime. }
  \label{fig:NumStd}
\end{figure*}
\section{ Numerical Results }\label{sec:Num_Stud}
In this section, we evaluate the performance of FL under different scheduling policies through both numerical analysis and experimental simulations. Specifically, we start with the numerical study to draw insights and then follow with simulations for validation.
Unless otherwise stated, the following system parameters will be used: AP deployment density $\lambda = 10^{-4} \mathrm{m}^2$, number of associated UEs per cell $K=100$, number of orthogonal subchannels $N=10$, path loss exponent $\alpha = 3.8$, and SINR decoding threshold $\theta = 0$ dB.
\subsection{ Numerical Study }
We first explore the effect of the network parameters on the convergence rate of FL using the analysis derived in Section~VI.
Because the value of the data set size $n$ and the targeted duality gap $\varepsilon$ often depend on specific tasks, we adopt a ``normalized'' performance metric by dividing the required communication rounds with respect to $\log(n/\varepsilon)$ and refer to this quantity as normalized communication rounds.

In Fig.~\ref{fig:NumStd}, we plot the normalized communication rounds as a function of the total group number, $G$, under two different SINR operating regimes, namely, high SINR ($\theta=15$ dB) and low SINR ($\theta=-25$ dB).
Fig.~\ref{fig:1a} reveals that under high SINR threshold, running FL with PF results in a large reduction in the iteration time compared to  those with RS and RR, whereas the latter two schemes have similar convergence performance. This observation is in line with Remark~\ref{rmk:High_SINR}, and the reason stems from the fact that high SINR threshold reduces the chance of successful transmission from an arbitrary UE, while PF improves the convergence rate by selecting UEs with better channel quality for the radio access so as to increase their transmission success probability.
On the other hand, it can be seen from Fig.~\ref{fig:1b} that for networks operating under low SINR threshold, RR outperforms both RS and PF, which is in line with Remark~\ref{rmk:Low_SINR}.
This is because in this scenario, transmissions from UEs can achieve very high success probability and the scheduling order becomes the bottleneck, i.e., guaranteeing the timely parameter update from each UE determines the convergence performance. Since RR is the fairest scheduling scheme among the three, it thus attains the best performance.
Fig.~\ref{fig:NumStd} also implies that the required iteration rounds increase almost linearly with respect to the total number of associated UEs, which coincides with the simulation result in \cite{NisYon:18}.
This is because additional UEs not only reduce the selection probability for radio access but also creates new updates that will be subjected to staleness, which together prolong the communication rounds.

\begin{figure}[t!]
  \centering

	{\includegraphics[width=0.95\columnwidth]{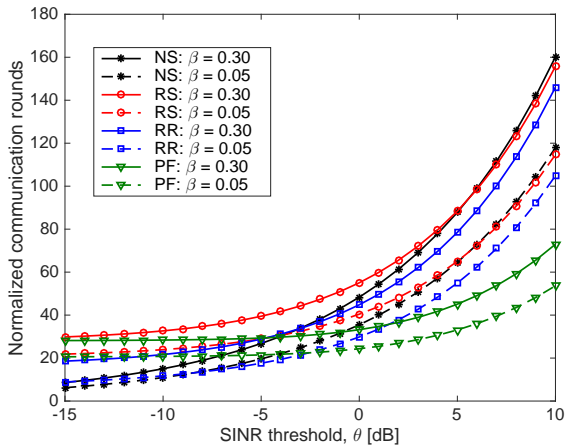}}
  \caption{ Normalized communication rounds vs SINR decoding threshold, where $N=10$ and $G=20$.  }
  \label{fig:NumCmp}
\end{figure}
Fig.~\ref{fig:NumCmp} further illustrates the normalized communication rounds as a function of the SINR decoding threshold, under different error levels.
This figure illustrates two phenomena.
One, in wireless networks with low SINR threshold, simply running FL without any scheduling, namely the no schedule (NS) approach, can outperform those with specific scheduling policies, because the impact of interference is minor and the success probability is high.
However, the performance of FL under NS quickly worsens as the SINR threshold goes up, while the ones with good scheduling schemes, e.g., the PF scheduling, are able to keep the required communication rounds at a low level.
Hence, adopting appropriate scheduling policies in different SINR regimes is critical to achieving good convergence performance of FL.
Second, regardless of the particular scheduling policy employed, the required communication rounds toward a given duality gap ratchets up as the SINR threshold increases.
As such, reducing the dimension of the updated parameters via compression or quantization \cite{NekAlpNai:18} so as to maintain a relatively small decoding threshold at the AP side is important to improve the convergence rate of FL in wireless networks. In fact, from Fig.~3 we can see that if quantization can achieve a 5~dB reduction in the decoding threshold, e.g., decreasing it from 10~dB to 5~dB, then even though that gives rise to a six-fold higher error level (namely $\beta$ grows from 0.05 to 0.3) the resulting convergence rate is nevertheless better than the original one. Further, if the decoding threshold can be reduced from 10~dB to 0~dB, then RR can achieve similar convergence rate as PF with conservative parameter compression, i.e., the one with 0.05 error level and decoding threshold of 10~dB.
To this end, the tradeoff between error level and decoding threshold is of importance to study.

\begin{figure}[t!]
  \centering

	{\includegraphics[width=0.95\columnwidth]{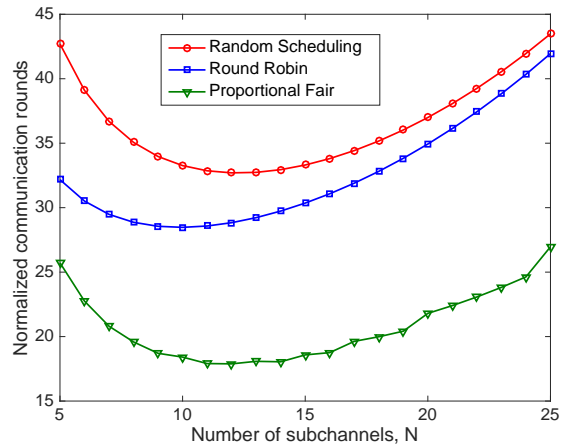}}
  \caption{ Normalized communication rounds vs number of subchannels, $N$.  }
  \label{fig:NumChnl}
\end{figure}
Fig.~\ref{fig:NumChnl} compares the normalized communication rounds of FL under RS, RR, and PF as a function of the number of subchannels, $N$. Note that an optimal $N$ that minimizes the required communication rounds exist for each of the scheduling policies, due to a trade-off between simultaneously serving more UEs and attaining higher success probability in each round of transmission.
The figure shows that in each update iteration, fewer UEs should be scheduled under RS and RR, thus leaving more spectrum to enhance the transmission success probability. On the other hand,
as PF is able to choose the UEs with good channel quality for the update, it thus allows more UEs to be selected while maintaining the transmissions success probability, which further accelerates the convergence rate of FL.

In summary, among the three scheduling policies, i.e., RS, RR, and PF, the PF has the best performance in scenarios with high SINR threshold, while RR is preferable when the SINR threshold is low. Moreover, the detection threshold has a direct impact on the required running time of the algorithm, thus quantizing weights into a lower dimension is more desirable in wireless FL. Further, there is a trade-off between the number of scheduled UEs and the subchannel bandwidth in optimizing the FL convergence rate, thus leaving room for further design opportunities.

\begin{figure}[t!]
  \centering

	{\includegraphics[width=0.95\columnwidth]{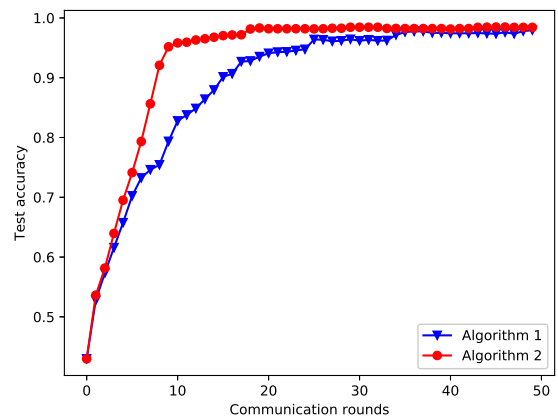}}
  \caption{ Comparison between Algorithm~1 and Algorithm~2: $K=100$, $N=10$, and $\theta = - 10$~dB.  }
  \label{fig:CmprAlg}
\end{figure}

\subsection{ Experimental Study }
\begin{figure*}[t!]
 \centering
   \subfigure[\label{fig:svm.a}]{\includegraphics[width=0.95\columnwidth]{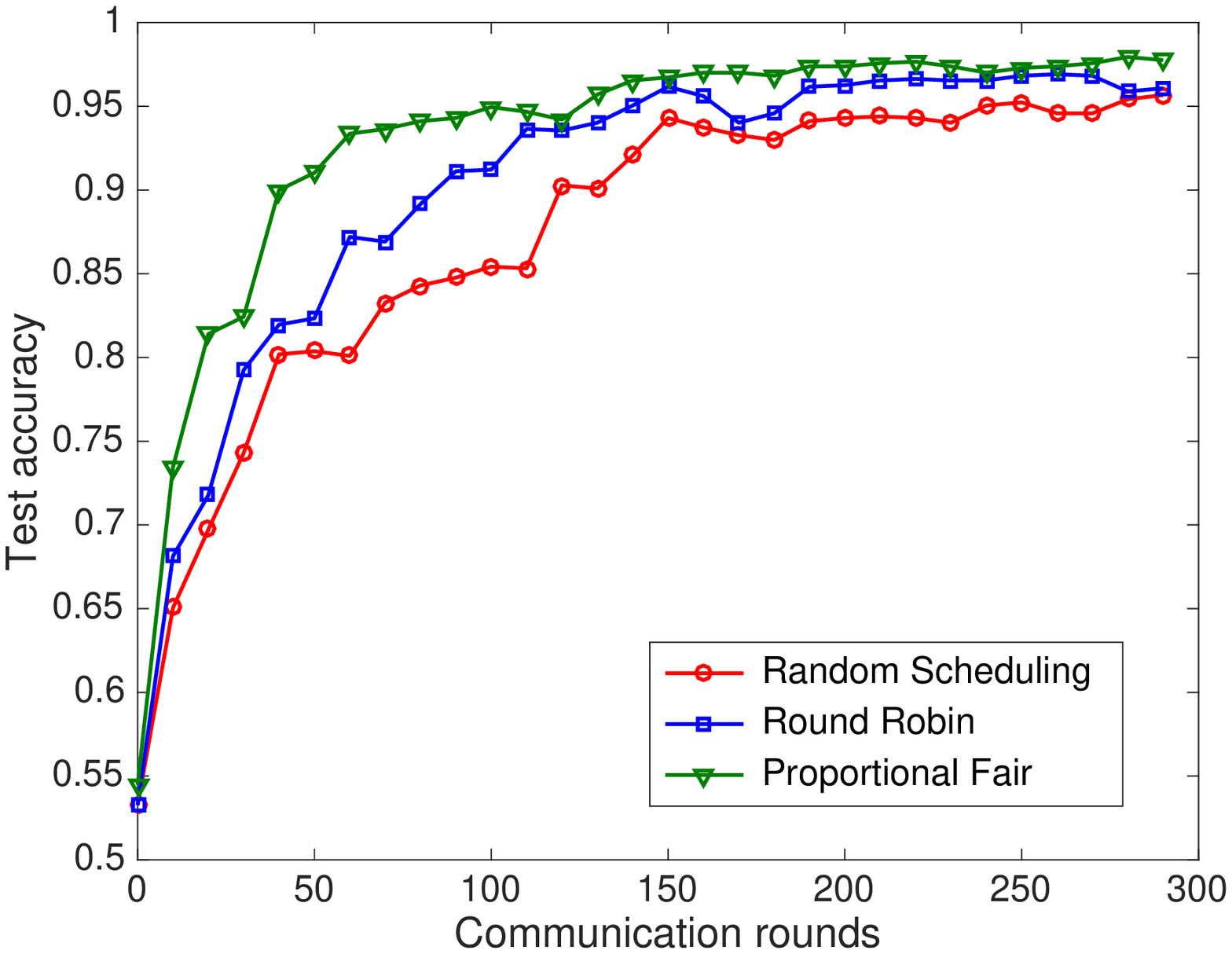}}~\qquad\quad
 \subfigure[\label{fig:svm.b}]{\includegraphics[width=0.95\columnwidth]{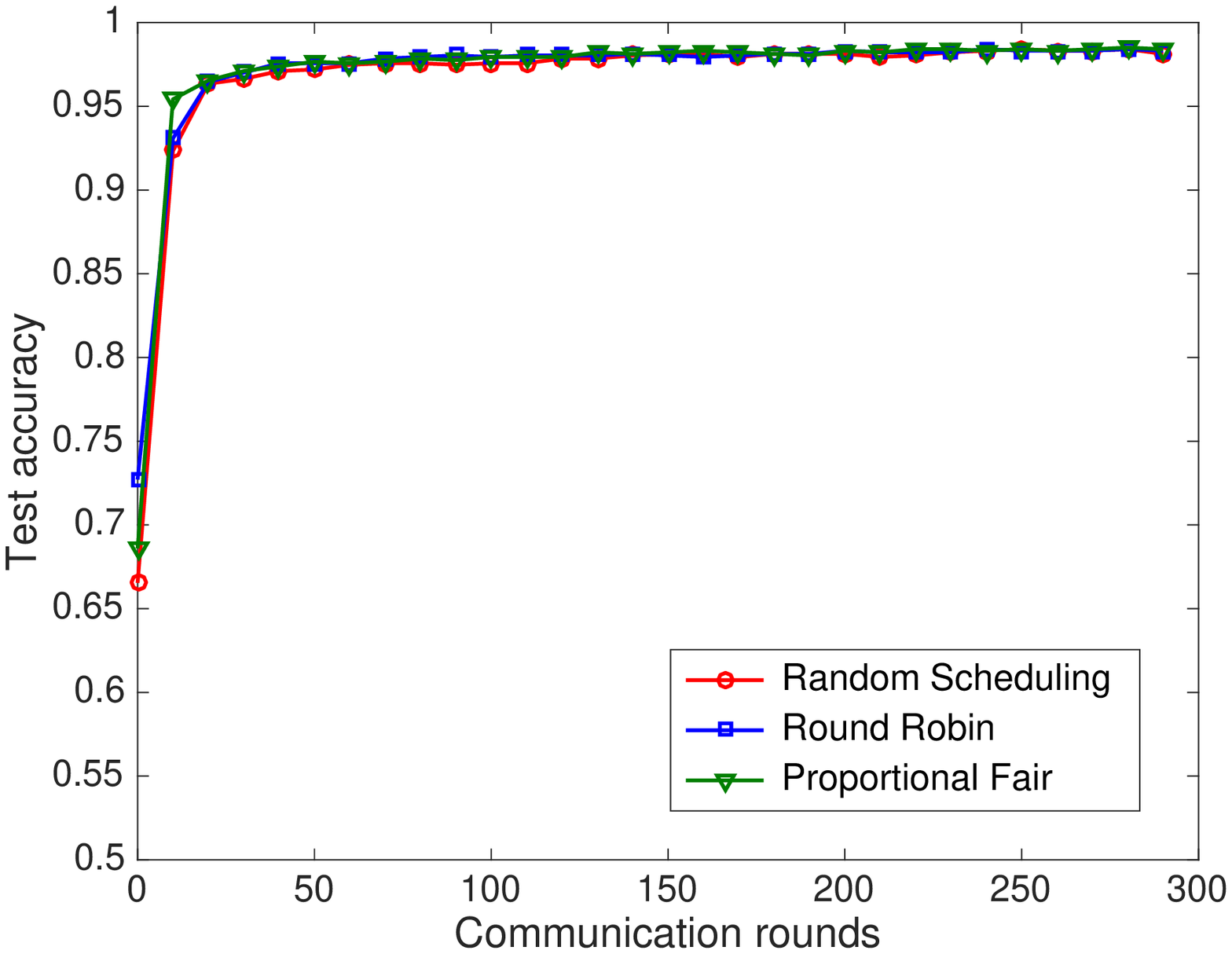}}
 \caption{ Test performance of the trained SVM with different scheduling policies RS, RR, and PF. The results are averaged over 20 trails under ($a$) high SINR threshold, $\theta = 20$~dB and ($b$) low SINR threshold, $\theta = -25$~dB. }
 \label{svm}
\end{figure*}
\begin{figure*}[t!]
 \centering
   \subfigure[\label{fig:svm.a}]{\includegraphics[width=0.95\columnwidth]{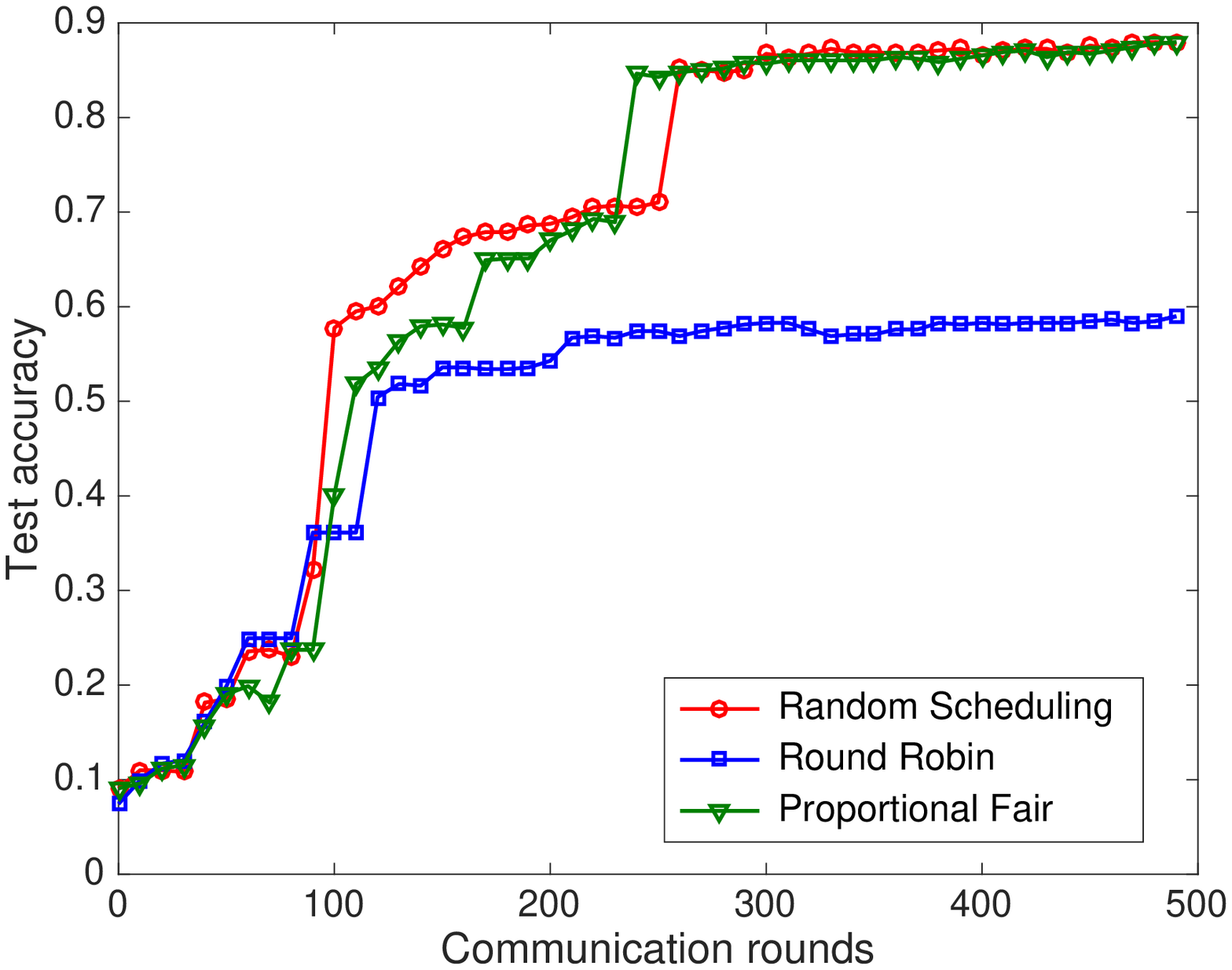}}~\qquad\quad
 \subfigure[\label{fig:svm.b}]{\includegraphics[width=0.95\columnwidth]{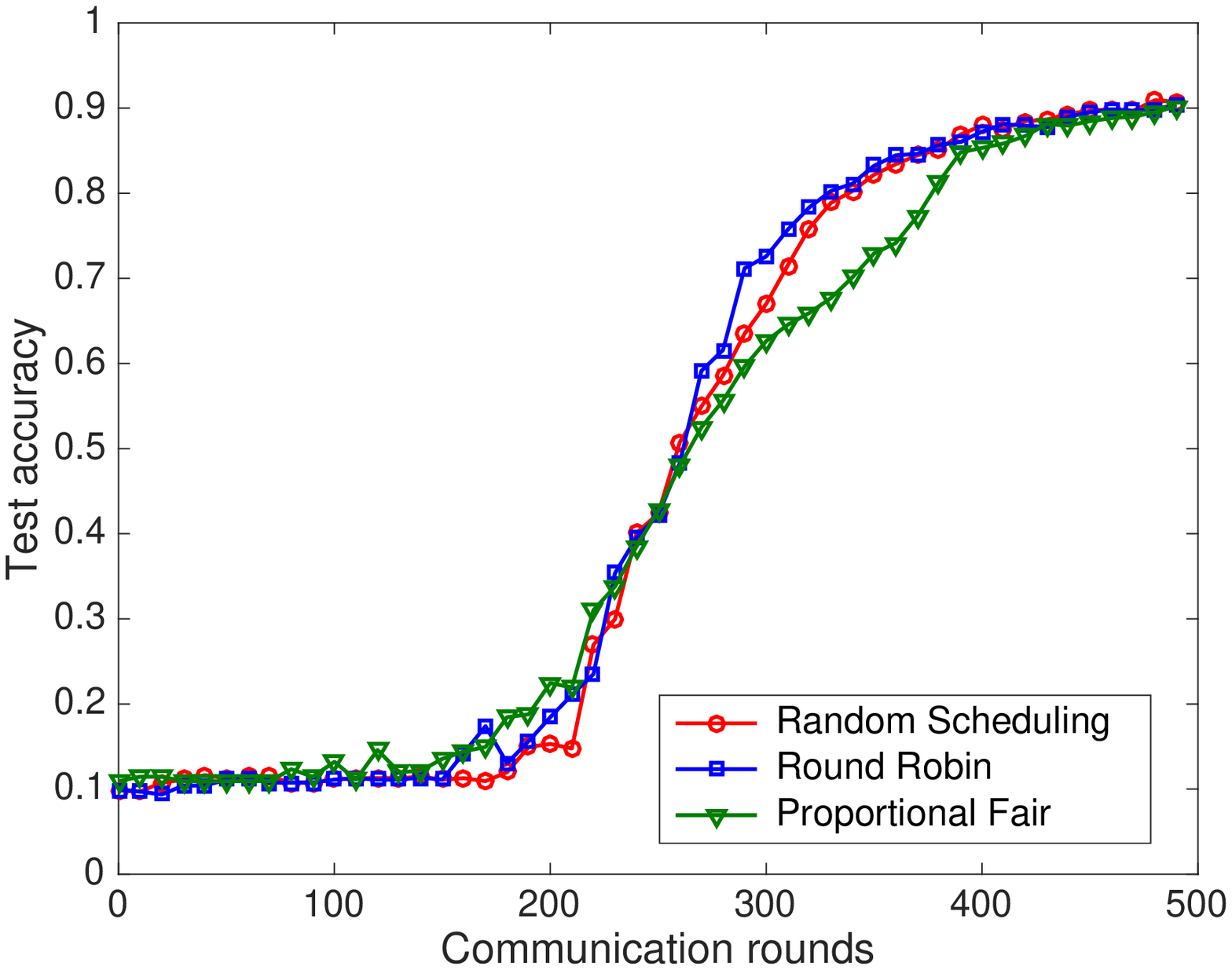}}
 \caption{ Test performance of the trained CNNs with different scheduling policies RS, RR, and PF. The results are averaged over 5 trials under ($a$) high SINR threshold, $\theta = 20$~dB and ($b$) low SINR threshold, $\theta = -25$~dB. We set K = 30 for the high SINR regime and K = 100 for the low SINR regime for a better illustration. }
 \label{cnn}
\end{figure*}

In this section, we showcase the effects of scheduling policies under different SINR scenarios. We first compare the performance between Algorithm~\ref{Alg:Gen_FL} and Algorithm~\ref{Alg:Wireless_FedSGD} in Fig.~\ref{fig:CmprAlg}, which plots the result of training an SVM on the MNIST data set, which consists of handwritten numerals. As figure shows, the proposed Algorithm~2 attains better convergence performance than the vanilla approach Algorithm~1. This mainly is due to the fact that Algorithm~2 can leverage advanced approaches to tackle the local subproblems than merely adopting the SGD.

Next, two machine learning models, namely an SVM and a convolutional neural network (CNN){\footnote{Note that a CNN has a non-convex objective function and hence the analysis of this paper does not directly apply to this model. Nonetheless, this experiment demonstrates that similar behavior may still hold under non-convex objective functions.}}, are evaluated by clamping the low SINR threshold as $\theta = -25$ dB and high SINR threshold as $\theta = 20$ dB. The number of UEs for each AP is $K = 100 \text{ or }  30$, and the number of subchannels is $N = 5$.
For the SVM, we consider a two-class classification task to recognize digits $0$ and $8$ where each UE is assigned with 5 training samples. We also evaluate the CNN for the multi-class classification task, namely, recognizing from $0$ to $9$, where each UE has 100 training samples locally. The models are tested every 10 training steps over 1000 test samples. Results are reported in Fig.~\ref{svm} and Fig.~\ref{cnn}. The learning rate is $\eta = 0.01$ for both models.

The results in the higher SINR regime are consistent with the theorems, i.e., PF theoretically converges faster than RR. We can observe from Fig.~\ref{svm}.(a) that at a higher SINR threshold, i.e., $\theta = 20$ dB, the SVM model trained with PF reaches a steady stage in 60 training steps while that trained with RR takes around 100 steps. Also note that RS is worse than RR or PF in this scenario. The advantage of PF over RR in the high SINR regime is even obvious for more complicated models such as CNNs. As shown in Fig.\ref{cnn}, models trained with PF achieve an average accuracy of 0.94 while models trained with RR get stuck in an accuracy of 0.5. This is mainly due to the fewer successful global aggregations in RR as opposed to PF where subchnnels with highest SINR are invariably selected. Note that RS also performs similarly to PF because of the relatively higher probability for successful aggregations when $K$ is small.

We also report the results in the low SINR regime, as shown in Fig.\ref{svm}.(b) and Fig.\ref{cnn}.(b). We notice that the performance gap among different scheduling policies  disappears when the model is very simple. For example, PF, RR and RS exhibit almost the same performance when $\theta = -25$ dB. This is because every local UE is able to achieve reasonable performance on the classification task. Were this to happen, models trained with PF, RR and RS are expected to behave similarly, since global aggregation would be very likely successful when the SINR threshold is as low as $-25$ dB. When the model becomes more intricate, models trained with RR perform better than PF, as shown in Fig.\ref{cnn}.(b), which is also in an agreement with the above theorems.

\section{ Conclusion }\label{sec:Conclusion}
In this paper, we have undertaken an analytical study of the effects of three practical scheduling policies, i.e., random scheduling (RS), round robin (RR), and proportional fair (PF), to the performance of federated learning (FL) in wireless networks.
We used a general model that accounts for scheduling schemes, inter-cell interference, and resource allocation between the radio access links and the training stage. Our analysis has shown that running FL with PF is able to achieve much smaller iteration time than RS and RR if the network is operating under a high SINR threshold, while RR is more preferable when the SINR threshold is low.
Moreover, the convergence rate of FL decreases rapidly as the SINR threshold increases, confirming the importance of compression and quantization of the update parameters.
Our analysis has also revealed a trade-off between the number of scheduled UEs and the subchannel bandwidth under a fixed amount of available spectrum, showing further design opportunities.

The framework provided in this paper allows one to explicitly characterize the interplay between model training and parameter update phases in general FL algorithms, where stragglers and transmission failure can be severe depending on the transmission protocol and scheduling policies employed.
More generally, our work helps to understand how the key features of a wireless network, i.e., fading, path loss, interference, and deployment strategy, affect the convergence rate of FL running in such a context.
This paper has considered the current state-of-the-art scheduling policies deployed in practice. More advanced scheduling policies that account for both RR and PF can be considered, and improving the FL performance via more advanced wireless technologies, e.g., massive multiple-input-multiple-output (MIMO), full-duplex transmissions, or nonorthogonal multiple access (NOMA) is also a concrete direction.

\begin{appendix}
\subsection{Proof of Lemma~\ref{lma:Convg_Bnd}} \label{apx:Convg_Bnd}
Without loss of generality, we assume the learning process has progressed to the $t$-th communication round. Upon completion, the global parameters will be updated from $\mathbf{v}(\mathbf{a}^t)$ to $\mathbf{v}(\mathbf{a}^{t+1})$, whereas the local parameters at UE $k$ are updated to $\mathbf{a}^t_{[k]} + \eta^t \Delta \mathbf{a}^t_{[k]}$.
On the one hand, according to the duality between smoothness and strong convexity, we know that given a closed convex function $f$, it holds that if $f$ is $x$-strongly convex (resp. smooth), the conjugate function $f^*$ is (1/$x$)-smooth (resp. strongly convex) \cite[Theorem 4.2.1, 4.2.2]{HirJeaLem:12}.
Hence, following Assumptions 1 and 2, we have that the functions $\ell^*(\cdot)$ are $\mu$-strongly convex and $r^*(\cdot)$ is $1/\zeta$-smooth. On the other hand, the update aggregation in (19) can be written as
\begin{align}
\mathbf{v}(\mathbf{a}^{t+1}) \!=\! \mathbf{v}(\mathbf{a}^t) + \sum_{k=1}^K \Delta \mathbf{v}_k \mathbbm{1}\{ \mathcal{S}^{z}_{k,t} \!=\! 1, \gamma_{k,t} > \theta \}.
\end{align}
As such, the expectation of the updated objective function \eqref{equ:Dual_FedSGD} can be written as
\begin{align} \label{equ:E_Dvtp1}
& \mathbb{E}\big[ D( \mathbf{a}^{t+1} ) \big] = \mathbb{E}\Big[ \sum_{ k=1 }^K  - R_k( \mathbf{a}^t_{ [k] } + \eta^t \Delta \mathbf{a}^t_{ [k] } )
\nonumber\\
& -\frac{ \xi }{ K } r^*\big(\, \mathbf{v}(\mathbf{a}^t) + \sum_{k=1}^K \Delta \mathbf{v}_k^t \!\cdot\! \mathbbm{1}\{ \mathcal{S}^{z}_{k,t} \!=\! 1, \gamma_{k,t} \!>\! \theta \} \big) \, \Big].
\end{align}
It can be seen that the right hand side of the above equation contains $K$ local terms and one global term. We can thus deal with them individually. First of all, we deal with the global update term. Because $r^*(\cdot)$ is $1/\zeta$-smooth, the following holds:
\begin{align}\label{equ:E_rtp1}
& \mathbb{E}\big[ \frac{\xi}{K} r^*\big( \mathbf{v}(\mathbf{a}^{t}) \!+\! \sum_{k=1}^K \Delta \mathbf{v}^t_k \!\cdot\! \mathbbm{1}\{ \mathcal{S}^{z}_k \!=\! 1, \gamma_{k,t} > \theta \} \big) \big]
\nonumber\\
 \leq &  \frac{\xi}{K} \Big \{ r^*(\mathbf{v}(\mathbf{a}^{t})) + \frac{ 1 }{ 2 \zeta } \, \mathbb{E}\Big[ \big\Vert \sum_{k=1}^K \mathbf{v}_k^t \mathbbm{1}\{ \mathcal{S}^{z}_k \!=\! 1, \gamma_{k,t} > \theta \} ) \big\Vert^2 \Big]
 \nonumber\\
 & + \sum_{ k=1 }^K \nabla r^*(\mathbf{v}(\mathbf{a}^{t}))^T \Delta \mathbf{v}_k^t \, \mathbb{E}\big[ \mathbbm{1}\{ \mathcal{S}^{z}_k \!=\! 1, \gamma_{k,t} \!>\! \theta \} \big]  \Big\}
\nonumber\\
\leq & \frac{1}{K} \Big\{ \xi r^*(\mathbf{v}(\mathbf{a}^{t})) + \mathcal{U}_k^z \frac{ \kappa / \xi }{ 2 n^2 }   \sum_{ k=1 }^K  \Vert \mathbf{X}_{ [k] } \Delta \mathbf{a}^t_{ [k] } \Vert^2
\nonumber\\
& \qquad \qquad \quad + \mathcal{U}_k^z \sum_{ k=1 }^K \langle \frac{1}{n} \mathbf{X}_{[k]}^T \nabla r^*\!( \mathbf{v}(\mathbf{a}^{t}) ) , \Delta \mathbf{a}^t_{[k]}  \rangle \Big\}.
\end{align}
On the other hand, as $\ell^*_i(\cdot)$ are $\mu$-strongly convex, it follows that $R_k(\cdot)$ are also $\mu$-strongly convex. Using the convexity of $R_k(\cdot)$, we have
\begin{align}\label{equ:Rk_tp1}
& \mathbb{E}\big[ R_k( \mathbf{a}^t_{ [k] } +  \eta^t  \Delta \mathbf{a}^t_{ [k] } ) \big]
\nonumber\\
 =\, & \mathbb{E} \big[ R_k\big( \, [\, 1 -  \eta^t \, ]\, \mathbf{a}^t_{ [k] } +  \eta^t  ( \mathbf{a}^t_{ [k] } + \Delta \mathbf{a}^t_{ [k] } )\, \big) \big]
\nonumber\\
\leq \, & \mathbb{E}\big[  ( 1 - \eta^t ) R_k( \mathbf{a}^t_{ [k] } ) + \eta^t R_k (  \mathbf{a}^t_{ [k] } + \Delta \mathbf{a}^t_{ [k] } ) \big]
\nonumber\\
=\, & ( 1 - \mathcal{U}_k^z ) R_k( \mathbf{a}^t_{ [k] } ) + \mathcal{U}_k^z R_k (  \mathbf{a}^t_{ [k] } + \Delta \mathbf{a}^t_{ [k] } ).
\end{align}
By substituting \eqref{equ:E_rtp1} and \eqref{equ:Rk_tp1} into \eqref{equ:E_Dvtp1}, we have
\begin{align}
&\mathbb{E}\big[ D( \mathbf{a}^{t+1} ) \big] \geq ( 1 - \mathcal{U}^z_k ) \Big\{ - \xi r^*(\mathbf{v}(\mathbf{a}^{t})) - \sum_{ k=1 }^K R_k( \mathbf{a}^t_{ [k] } ) \Big\}
\nonumber\\
& + \mathcal{U}^z_k \Big\{ - \xi r^*( \mathbf{v}(\mathbf{a}^{t}) )  - \sum_{ k = 1 }^K \nabla r^*(\mathbf{v}(\mathbf{a}^{t}))^T \Delta \mathbf{v}^t_{k}
\nonumber\\
& - \frac{ \kappa / \xi }{ 2 n^2 } \sum_{ k=1 }^K  \big\Vert \mathbf{X}_{ [k] } \Delta \mathbf{a}^t_{ [k] } \big\Vert^2 - \sum_{ k = 1 }^K R_k (  \mathbf{a}^t_{ [k] } + \Delta \mathbf{a}^t_{ [k] } ) \Big\},
\end{align}
and the result follows by substituting \eqref{equ:D_a} and \eqref{equ:Local_DelSuP} into the above inequality.

\subsection{Proof of Theorem~\ref{thm:CnvRate_Genl}} \label{apx:CnvRate_Genl}
After receiving updates from the $t$-th to the $(t+1)$-th communication round, the expected increment in the objective function is
\begin{align}
& \mathbb{E}\Big[ D( \mathbf{a}^{ t+1 } ) - D( \mathbf{a}^{ t } ) \Big]
\nonumber\\
\geq \, & \, \mathcal{U}^{ z }_k \Big[ \sum_{ k=1 }^K \Delta D( \Delta \mathbf{a}^*_{ [k] }; \mathbf{v}(\mathbf{a}^{t}), \mathbf{a}^t_{ [k] } ) - \mathbb{E} \big[D( \mathbf{a}^{t} ) \big]
\nonumber\\
&  \!+\! \sum_{ k=1 }^K \! \Delta D( \Delta \mathbf{a}^t_{ [k] }; \mathbf{v}(\mathbf{a}^{t}), \mathbf{a}^t_{ [k] } ) \!-\!\! \sum_{ k=1 }^K \!\! \Delta D( \Delta \mathbf{a}^*_{ [k] }; \mathbf{v}(\mathbf{a}^{t}), \mathbf{a}^t_{ [k] } ) \Big]
\nonumber\\
\stackrel{(a)}{\geq} & (1\!-\!\beta) \, \mathcal{U}^{ z }_k  \Big[ \sum_{ k=1 }^K \Delta D( \Delta \mathbf{a}^*_{ [k] }; \mathbf{v}(\mathbf{a}^{ t }), \mathbf{a}^t_{ [k] } ) - \mathbb{E}\big[ D( \mathbf{a}^{ t } ) \big] \Big]
\end{align}
where $(a)$ follows from Assumption~\ref{appm:prec_levl} and noticing that $D( \mathbf{a}^{t} )=\sum_{ k=1 }^K \Delta D( \mathbf{0}; \mathbf{v}(\mathbf{a}^{t}), \mathbf{a}^t_{ [k] } )$. Moreover, because $\ell_i(\cdot)$ is $1/\mu$-smooth, $\ell_i^*(\cdot)$ is $\mu$-strongly convex. Hence, there exist a scalar $s \in [0,1]$ and an $n$-dimension vector $\mathbf{u}=(u_1, \cdots, u_n)$ whereas $\mathbf{u}_{[k]} \in \partial( R_k ) $ with $\partial(R_k)$ being the subgradient of $R_k$, such that $\Delta \mathbf{a}_{[k]}^t = s( \mathbf{u}_{[k]}^t - \mathbf{a}_{[k]}^t )$ and the following holds \cite{MaKonJag:17}:
\begin{align}
& \mathbb{E}\big[ D( \mathbf{a}^{t} ) \big]  - \sum_{ k=1 }^K \Delta D( \Delta \mathbf{a}^*_{ [k] }; \mathbf{v}(\mathbf{a}^{t}), \mathbf{a}^t_{ [k] } )
\nonumber\\
& \leq  \frac{1}{n} \sum_{i=1}^n \Big( - s \ell^*_i( - u_i ) - s \ell^*_i( - a_i ) - \frac{ \mu }{ 2 } (1-s)s (u_i - a_i)^2 \Big)
\nonumber\\
& + \! \langle \frac{1}{n} \mathbf{X}_{[k]}^T \nabla r^*\!( \mathbf{v}(\mathbf{a}^{t}) ) , \Delta \mathbf{a}^t_{[k]}  \rangle +\! \sum_{k=1}^K \frac{ \kappa / \xi }{ 2 n^2 } \big\Vert \mathbf{X}_{[k]} s ( \mathbf{u}_{[k]}^t \!-\! \mathbf{a}_{[k]}^t ) \big\Vert^2
\nonumber\\
& \leq \bar{s} \Big[ D( \mathbf{a}^* ) - D( \mathbf{a}^t ) \Big],
\end{align}
where $\bar{s} \in (0,1) $. As such, we have the following:
\begin{align} \label{equ:ConvRelat}
& \mathbb{E}\big[ D( \mathbf{a}^{*} ) - D( \mathbf{a}^{t+1} ) \big] \leq \mathbb{E}\big[  D( \mathbf{a}^{*} ) - D( \mathbf{a}^{t} ) \big]
\nonumber\\
 + & (1-\beta) \, \mathcal{U}^{ z }_k \, \big[ D( \mathbf{a}^{t} ) -  \sum_{ k=1 }^K \Delta D( \Delta \mathbf{a}^*_{ [k] }; \mathbf{v}(\mathbf{a}^{t}), \mathbf{a}^t_{ [k] } ) \big]
\nonumber\\
\leq & \big[ 1 - (1-\beta) \, \mathcal{U}^{ z }_k \big] \mathbb{E}\big[ D( \mathbf{a}^{*} ) - D( \mathbf{a}^{t} ) \big]
\nonumber\\
\leq & \big[ 1 - (1-\beta) \, \mathcal{U}^{ z }_k \big]^t \mathbb{E}\big[ D( \mathbf{a}^{*} ) - D( \mathbf{a}^{0} ) \big].
\end{align}
The result then follows by upper bounding the R.H.S. of \eqref{equ:ConvRelat} by $\varepsilon$ and noticing that $\mathbb{E}[ D( \mathbf{v}^{0} ) - D( \mathbf{v}^{*} )] < n$ \cite{SmiForMa:18}.

\subsection{Proof of Corollary~\ref{Cor:Cnv_RS}} \label{apx:Cnv_RS}
Using the law of total probability, the parameter update success probability of UE $k$ can be written as follows:
\begin{align}
\mathcal{U}^{ \mathrm{RS} }_{k} &= \mathbb{P}( \gamma_{k,t} > \theta, \mathcal{S}^{ \mathrm{RS} }_{k,t} = 1 )
\nonumber\\
&= P_{\mathrm{s}}( \gamma_{k,t} > \theta | \mathcal{S}^{ \mathrm{RS} }_{k,t} = 1 ) \mathbb{P}( \mathcal{S}^{ \mathrm{RS} }_{k,t} = 1 ).
\end{align}
For a generic UE, the probability of being selected by the AP for parameter update during one typical iteration is given by
\begin{align}
&\mathbb{P}( \mathcal{S}_{k,t}=1 ) = 1 - \mathbb{P}( \mathcal{S}_{k,t} = 0 )
\nonumber\\
= & 1 - \frac{ K-1 }{ K } \times \frac{ K-2 }{ K - 1 } \times \cdots \times \frac{ K-N }{ K - (N+1) }  = \frac{1}{G}.
\end{align}
Once UE $k$ is selected, the probability that its parameters can be successfully updated at the AP is equivalent to the probability that the received SINR exceeds the decoding threshold. Using tools from stochastic geometry \cite{BacBla:09}, we first condition on the distance $\Vert x_k \Vert = r_k$ and arrive at the following:{\footnote{The actual locations of uplink UEs form a Poisson-Voronoi perturbed lattice, and an exact interference characterization for this point process is not yet available. We thus approximate the locations by a non-homogeneous PPP \cite{Hae:17}, which gives a very tight approximation.}}
\begin{align} \label{equ:SucProb_Cond_RS}
& \mathbb{P}( \gamma_{k,t} > \theta | r_k, \mathcal{S}_{k,t} = 1 ) = \mathbb{P}\Big( h_{c_k} > \theta r_k^\alpha \big(\! \sum_{ c \in \tilde{\Phi}^k_{ \mathrm{u} } } \! \frac{ h_c }{ \Vert c \Vert^\alpha } \!+\! \frac{ \sigma^2 }{ \PVT } \big) \Big)
\nonumber\\
&\approx \mathbb{E}\Big[ \exp\!\big(\! -\! \frac{ \theta \sigma^2 r_k^\alpha }{ \PVT } \big) \exp\!\Big(\! -\! \lambda \pi \!\! \int_0^\infty\! \frac{ ( 1 - e^{ - \frac{12}{5} \lambda \Vert x \Vert^2 } ) }{ 1 + \Vert x \Vert^\alpha / \theta r_k^\alpha } dx \Big) \Big]
\nonumber\\
&= \exp\!\Big(  \frac{ - \theta \sigma^2 r_k^2 }{ \PVT (2 \lambda)^{ \frac{ \alpha }{ 2 } - 1 } } -\! \lambda \pi r_k^2 \theta^{ \frac{ 2 }{ \alpha } } \!\!\! \int_0^\infty \! \frac{ 1 - e^{ - \frac{12}{ 5 \pi } \theta^{ \frac{2}{\alpha} } u} }{ 1 + u^{ \frac{\alpha}{2} } } du \Big).
\end{align}
Notice that the probability density function of $r_k$ follows Rayleigh distribution \cite{BacBla:09}
\begin{align} \label{equ:Dist_R0}
f_{R_k}(r_k) = 2 \pi \lambda r_k \exp( - \lambda \pi r_k^2 );
\end{align}
we can thus decondition $r_k$ in \eqref{equ:SucProb_Cond_RS} according to \eqref{equ:Dist_R0} and obtain the desired result.

\subsection{Proof of Corollary~\ref{Cor:Cnv_RR}} \label{apx:Cnv_RR}
By employing RR, each UE is selected to transmit per $G$ communication rounds. As such, the selected probability of a typical UE is
  \begin{align}\label{equ:ActProb_Gnrl}
  \mathbb{P}( \mathcal{S}_{k,t}^{ \mathrm{RR} } = 1 ) = \left \{
  \begin{tabular}{cc}
  \!\!\!\! $1$, & $\mathrm{if~scheduled}$,   \\
  \!\!\!\!  $0$, & $\mathrm{otherwise}$.
  \end{tabular}
  \right.
  \end{align}
The result then follows by noticing that $\mathbb{P}( \gamma_{k,t} > \theta | \mathcal{S}_{k,t}^{ \mathrm{RR} } = 1 )$ can be calculated via the same approach as in RS.
Under RR, the trained parameter from any particular UE is updated once per $G$ communication rounds. Without loss of generality, we assume that the update of UE $k$ starts at time index $0$. As such, each communication epoch of UE $k$ occurs at $t=mG$, $m \in \mathbb{N}$. Thus, using Theorem~\ref{thm:CnvRate_Genl} and similar approach as in \eqref{equ:ConvRelat}, we have
\begin{align} \label{equ:RR_Bound}
& \mathbb{E}\big[ D( \mathbf{a}^{*} ) - D( \mathbf{a}^{t+1} ) \big] \leq \mathbb{E}\big[  D( \mathbf{a}^{*} ) - D( \mathbf{a}^{t} ) \big]
\nonumber\\
\leq & \big[ 1 - (1-\beta) \, \mathcal{U}^{ \mathrm{RR} }_k \big]^t \mathbb{E}\big[ D( \mathbf{a}^{*} ) - D( \mathbf{a}^{0} ) \big]
\nonumber\\
= & \Big( 1 - \frac{ 1-\beta }{ 1 + \mathcal{V}(\theta, \alpha) } \Big)^{ \lfloor \frac{t}{G} \rfloor  } \mathbb{E}\big[ D( \mathbf{a}^{*} ) - D( \mathbf{a}^{0} ) \big]
\end{align}
where $\lfloor \cdot \rfloor$ is the floor function.
By upper bounding \eqref{equ:RR_Bound} by $\varepsilon$, we arrive at the desired result.

\subsection{Proof of Corollary~\ref{Cor:Cnv_PF}} \label{apx:Cnv_PF}
Due to the stationary property of PPPs, in the steady state, the average SNR from each UE will be indentical and the PF is equivalent to selecting $N$ UEs out of $K$ with the highest channel gains \cite{ChoBah:07}. As such, a typical UE $k$ will be selected only if its channel gain is among the highest $N$ out of the $K$ UEs, i.e.,
\begin{align}\label{equ:Slc_PF}
\mathbb{P}( \mathcal{S}_{k,t} = 1 ) = \frac{ N }{ K } = \frac{ 1 }{ G }.
\end{align}
And the channel gain of the selected UE can be written as $h_k = \max\{ h_{i_1}, ..., h_{ i_{K-N+1} } \}$ which has the following distribution:
\begin{align}
\mathbb{P}( h_k < h ) = \!\! \prod_{ m=1 }^{K-N+1} \!\! \mathbb{P}( h_{i_m} < h ) = ( 1 - e^{-h} )^{K - N + 1}.
\end{align}
The transmission success probability of a selected UE can then be calculated as
\begin{align}\label{equ:Suc_PF}
& \mathbb{P}( \gamma_{k,t} \!>\! \theta ) = 1- \mathbb{E}\bigg[ \! \Big( 1 \!-\! e^{ - \theta \Vert c_k \Vert^\alpha \! \big( \! \sum_{ c \in \tilde{\Phi}^k_{ \mathrm{u} } } \! \frac{ h_c }{ \Vert c \Vert^\alpha } \!+\! \frac{ \sigma^2 }{ \PVT } \big)} \Big)^{ N \!-\! K \!+\! 1 } \bigg]
\nonumber\\
&=\!\! \sum_{ i=1 }^{ K\!-\!N\!+\!1 } \!\!\! \binom{ K \!-\! N \!+\! 1 }{ i } (-1)^{i+1} \mathbb{E}\bigg[ e^{ - i \theta \Vert c_k \Vert^\alpha \! \big( \! \sum_{ c \in \tilde{\Phi}^k_{ \mathrm{u} } } \! \frac{ h_c }{ \Vert c \Vert^\alpha } \!+\! \frac{ \sigma^2 }{ \PVT } \big)} \bigg]
\nonumber\\
&\stackrel{(a)}{=} \sum_{ i=1 }^{ K - N + 1 } \! \binom{ K - N + 1 }{ i } \frac{ (-1)^{i+1} }{ 1 + \mathcal{V}(i\theta, \alpha) },
\end{align}
where the derivation in ($a$) follows a similar approach as that in the proof of Corollary~\ref{Cor:Cnv_RS}.
We obtain the result by taking \eqref{equ:Slc_PF} and \eqref{equ:Suc_PF} to compute the parameter update success probability.

\end{appendix}
\bibliographystyle{IEEEtran}
\bibliography{bib/StringDefinitions,bib/IEEEabrv,bib/howard_FedSGD}

\end{document}